\documentclass[paper]{JHEP3}

\usepackage{epsfig}
\usepackage{amssymb}
\usepackage{feynarts}
\def\URLtilde{\lower0.2em\hbox{$\tilde{\phantom{a}}$}}

\def\beq{\begin{equation}}
\def\eeq{\end{equation}}

\def\beqn{\begin{eqnarray}}
\def\eeqn{\end{eqnarray}}

\def\mhat{\widehat{m}}
\def\red{
}
\def\black{
}
\def\mycomm#1{\hfill\break\strut\kern-3em{\red\tt ====> #1\black}\hfill\break}

\newcommand{\ud}{\mathrm{d}}

\newcommand{\llp}{\mathcal{M}_{ll}^+}
\def\mqq{m^2_D}
\def\mqr{m^4_D}
\def\mqn{m^2_{qn}}
\def\mqm{m^4_{qn}}
\def\mqf{m^2_{qf}}
\def\mll{m^2_{ll}}

\preprint{Cavendish--HEP--06/11(rev.)}
\title{Distinguishing Spins in Decay Chains at the Large Hadron Collider%
\footnote{Work supported in part by the UK Particle Physics and
Astronomy Research Council.}}
\author{Christiana Athanasiou$^1$, Christopher G.\ Lester$^2$,
Jennifer M.\ Smillie$^3$ and Bryan R.\ Webber$^4$\\
  Cavendish Laboratory, University of Cambridge,\\
  JJ Thomson Avenue, Cambridge CB3 0HE, U.K.\\
  $^1$E-mail: \email{c.athanasiou.02@cantab.net}\\
  $^2$E-mail: \email{lester@hep.phy.cam.ac.uk}\\
  $^3$E-mail: \email{smillie@hep.phy.cam.ac.uk}\\
  $^4$E-mail: \email{webber@hep.phy.cam.ac.uk}}
\abstract{If new particles are discovered at the LHC, it will be
important to determine their spins in as model-independent a way
as possible. We consider the case, commonly encountered in models
of physics beyond the Standard Model, of a new scalar or fermion
$D$ decaying sequentially into other new particles $C,B,A$ via
the decay chain $D\to Cq$, $C\to Bl^{\rm near}$, $B\to Al^{\rm far}$,
$l^{\rm near}$ and $l^{\rm far}$ being opposite-sign same-flavour
charged leptons and $A$ being invisible. We compute the observable
2- and 3-particle invariant mass distributions for all possible spin
assignments of the new particles, and discuss their distinguishability
using a quantitative measure known as the Kullback-Leibler distance.}
 \keywords{Hadronic Colliders, Beyond Standard Model,
Supersymmetry Phenomenology, Large Extra Dimensions}

\begin{document}



\section{Introduction}
\label{sec:introduction}

The discovery of new physics at the TeV scale will be a principal objective of experiments
at the soon-to-be completed Large Hadron Collider (LHC).  In most scenarios for physics
beyond the Standard Model (BSM), new strongly-interacting particles will be observed if
the collision energy and luminosity are sufficiently high.  Typically these particles are
expected to decay weakly into cascades of Standard Model particles and, possibly, a stable
or metastable lightest new particle. In supersymmetry (SUSY) with R-parity, for example,
produced squarks will decay into quarks and, depending on the mass spectrum, leptons
and/or vector or Higgs bosons and the lightest supersymmetric particle, most often an
unobservable neutralino.

Until recently, discussions of new physics at the LHC tended to concentrate on how to
determine the free parameters of a particular, usually supersymmetric, model. Now that the
completion of the LHC is near, however, there is increasing interest in more
model-independent approaches.  Part of this interest arises from the realisation that
there are BSM scenarios in which the spins of produced particles differ from those
expected on the basis of supersymmetry.  Therefore one needs to consider more generally
the ways in which the spins of new particles can be determined from their decay chains.
 
In the present paper we assume that a particular chain of decays that is common in various
models, starting from a new scalar or fermionic quark, has been identified and that all
the masses of the new particles involved in it are known.  We then study the extent to
which decay correlations, manifest in the invariant mass distributions of combinations of
observable decay products, would enable one to distinguish between the different possible
spin assignments of the new particles.  This paper is thus an extension of earlier work in
which the SUSY decay correlations were compared with uncorrelated phase
space~\cite{Barr:2004ze,Goto:2004cp} or with those of a model that has universal extra
dimensions (UED)\footnote{The similarities between SUSY and UED at hadron colliders were
  first pointed out in
  ref.~\cite{Cheng:2002ab}.}~\cite{Smillie:2005ar,Battaglia:2005zf,Datta:2005zs,Datta:2005vx,Barr:2005dz,Alves:2006df}.

In the following section we present the decay chains to be considered,
and the possible assignments of new particle spins and observed
particle chiralities.  In section~\ref{sec:spin-correlations} we present our
new results on the corresponding decay correlations. We derive simple
analytical formulae for the correlation coefficients in terms of the
masses in the decay chain, which should be of general use whatever the
mass spectrum might turn out to be.  These extend those already given in
\cite{Smillie:2005ar,Miller:2005zp}.  We show graphical results for two 
specific mass scenarios, one considered more probable in SUSY and
the other in a UED model. In both mass scenarios we compare the
correlations predicted by all the possible spin assignments.

As was emphasised by Barr~\cite{Barr:2004ze}, the observability
of some of the correlations depends on the fact that
the LHC is a proton-proton collider, so that scalar or fermionic
quarks are produced somewhat more copiously than their antiparticles.
For the purposes of illustration, we take the quark fraction to be
$f_q=0.7$, as was the case for the SUSY and UED models studied in
ref.~\cite{Smillie:2005ar}.

In section~\ref{sec:model-discrimination} we consider the
extent to which the different spin assignments can be distinguished
on the basis of a quantitative measure known as the Kullback-Leibler
distance~\cite{Kullback:1951}. This allows us to establish a lower
limit on the number of events needed to discriminate between any
two spin assignments at a given level of confidence. Experimental
effects such as resolution and backgrounds will of course increase
the number of events needed. However, these are dependent on details
of the detector and analysis, and so we do not consider them here.
The lower limits we compute, being independent of such details,
can be used to assess whether or not discrimination between two
particular spin assignments is possible even in principle
with a given amount of data.

We perform analyses of this type on all the observable invariant
mass distributions separately, and also a combined analysis of
the full three-dimensional phase space distribution.
Our results and conclusions are summarized in section~\ref{sec:conc}.
The more lengthy formulae are consigned to the appendices.

\clearpage

\section{The decay chain}
\label{sec:decay-chain}

\begin{figure}[!h]
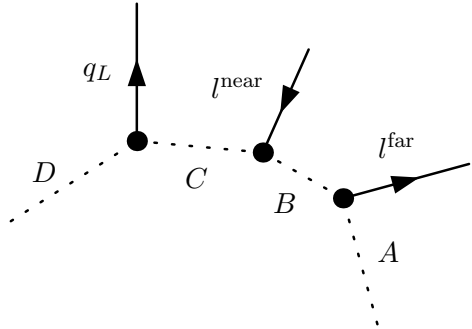

  \unitlength=0.5bp%
  \centering
  \begin{feynartspicture}(432,380)(1,1)    
    \FADiagram{}
    \FAProp(0.,10.)(5.5,13.5)(0.,){/GhostDash}{0}
    \FALabel(2.,11.7)[br]{$D$}
    \FAProp(5.5,19.5)(5.5,13.5)(0.,){/Straight}{-1}
    \FALabel(4.43,16.5)[r]{$q_L$}
    \FAProp(13.,17.5)(11.,13.)(0.,){/Straight}{1}
    \FALabel(11.0636,15.3995)[br]{$l^{\rm near}$}
    \FAProp(20.,12.5)(14.5,11.)(0.,){/Straight}{-1}
    \FALabel(16.8422,12.7654)[b]{$l^{\rm far}$}
    \FAProp(16.,5.5)(14.5,11.)(0.,){/GhostDash}{0}
    \FALabel(16.,8.65783)[l]{$A$}
    \FAProp(5.5,13.5)(11.,13.)(0.,){/GhostDash}{0}
    \FALabel(8.10967,12.3)[t]{$C$}
    \FAProp(11.,13.)(14.5,11.)(0.,){/GhostDash}{0}
    \FALabel(12.461,11.2)[tr]{$B$}
    \FAVert(5.5,13.5){0}
    \FAVert(11.,13.){0}
    \FAVert(14.5,11.){0}
  \end{feynartspicture}
  \vspace{-2cm}
  \caption{The decay chain under consideration.}
  \label{fig:decaychain}
\end{figure}

We will consider the cascade decay of a heavy colour-triplet scalar or fermion $D$, of the
form $D\to Cq,C\to Bl^{\rm near},B\to Al^{\rm far}$ (figure 1).  The decay products are
fixed as being a quark jet, a pair of opposite-sign-same-flavour (OSSF) leptons, and a
stable or long-lived massive new particle $A$.  We assume that the masses of the four
unknown particles $A$, $B$, $C$ and $D$ have already been measured (see
\cite{Allanach:2000kt,Lester:2001zx} for example, where edge analysis is used).  All
possible spin configurations in the decay chain are listed in table~\ref{tab:spins}.

\begin{table}[!ht]
  \centering
  \begin{tabular}{c | c | c | c}
    $D$ & $C$ & $B$ & $A$ \\ \hline 
    Scalar & Fermion & Scalar & Fermion \\ Fermion & Vector & Fermion & Vector \\
    Fermion & Scalar & Fermion & Scalar \\ Fermion & Vector & Fermion & Scalar \\
    Fermion & Scalar & Fermion & Vector \\ Scalar & Fermion & Vector & Fermion \\
  \end{tabular}
  \caption{Possible spin configurations in the decay chain (figure~\ref{fig:decaychain}).}
  \label{tab:spins}
\end{table}

These 6 chains will be labelled SFSF, FVFV, FSFS, FVFS, FSFV and SFVF respectively in what
follows.  Note that SFSF and FVFV are the SUSY and UED cases.

For fixed spin assignment, there are two possible angular distributions within the chain
as the quark and near lepton can have either the same or opposite helicity.  We will
follow the conventions of \cite{Barr:2004ze} and label these
\begin{itemize}
\item Process 1: $\{q,l^{\rm near},l^{\rm far}\} = \{q_L,l_L^-,l_L^+\}$ or
  $\{\bar{q}_L,l_L^+,l_L^-\}$ or $\{q_L,l_R^+,l_R^-\}$ or $\{\bar{q}_L,l_R^-,l_R^+\}$;
\item Process 2: $\{q,l^{\rm near},l^{\rm far}\} = \{q_L,l_L^+,l_L^-\}$ or
  $\{\bar{q}_L,l_L^-,l_L^+\}$ or $\{q_L,l_R^-,l_R^+\}$ or $\{\bar{q}_L,l_R^+,l_R^-\}$.
\end{itemize}
Note that in some of the processes above (FSFS and FSFV), spin information between the
quark and near lepton is lost as they are joined by a scalar.  For these chains,
processes 1 and 2 give the same distributions.

Treating the propagators of the unstable particles in the zero-width approximation
and neglecting all Standard Model particle masses, we can express the matrix elements
for these processes in terms of the mass of $A,B,C,D$ and the three two-particle
invariant masses of the quark plus near lepton, the quark plus far lepton, and the dilepton.
It will be convenient, as in \cite{Smillie:2005ar}, to introduce the mass ratios
\beq
\label{eq:xyz}
x= m_C^2/m_D^2\;,\qquad y= m_B^2/m_C^2\;,\qquad z= m_A^2/m_B^2\;,
\eeq
so that $0\le x,y,z \le 1$. The resulting formulae for
the full spin correlations are given in appendix~\ref{sec:MEs}.

To distinguish between the spin assignments, we integrate over two of the independent
variables and compare the predictions for the observable invariant mass distributions.
Throughout, we will show graphical results for two mass spectra of the
unknown particles $A,B,C,D$. The first (I) is the MSSM Snowmass point
SPS 1a \cite{Allanach:2002nj}, with fairly widely-spaced masses typical of SUSY.  The
relevant particles and their masses at this point are listed in table \ref{tab:SPS1a}. 
\begin{table}[!htbp]
  \centering
  \begin{tabular}{|c|c|c|c|}
    \hline $A$ & $B$ & $C$ & $D$ \\ 
    \hline $\tilde{\chi}_1^0$ & $\tilde{e}_R$ & $\tilde{\chi}_2^0$ & $\tilde{u}_L$ \\ 
    \hline 96 & 143 & 177 & 537 \\ \hline
  \end{tabular}
  \caption{Mass Spectrum I in GeV: Snowmass point SPS 1a.}
  \label{tab:SPS1a}
\end{table}

The second mass spectrum (II), with more nearly degenerate masses considered more likely
in a UED type scenario, is given in table \ref{tab:UED800}, where now the particles
involved are Kaluza-Klein excitations of Standard Model particles.  This UED spectrum was
calculated using the formulae for the radiative corrections given in \cite{Cheng:2002iz}
with $R^{-1} = 800$ GeV and $\Lambda R=20$.  Notice that in this scenario particle $C$
decays into left-handed leptons, whereas spectrum I involves right-handed leptons, as is
the case for MSSM point SPS 1a.

\begin{table}[!htbp]
  \centering
  \begin{tabular}{|c|c|c|c|}
    \hline $A$ & $B$ & $C$ & $D$ \\
    \hline $\gamma^*$ &
    $l_L^*$ & $Z^*$ & $q_L^*$ \\ \hline 800 & 824 & 851 & 956 \\ \hline
  \end{tabular}
  \caption{Mass Spectrum II in GeV: Calculated in UED with $R^{-1}=800$ GeV.}
  \label{tab:UED800}
\end{table}


\section{Invariant mass distributions}
\label{sec:spin-correlations}

\subsection{Dilepton mass distributions}
\label{sec:dilept-mass-distr}
The dilepton mass, $m_{ll}$, is the same in processes 1 and 2 and is also relatively easy
to measure, making it a potentially powerful tool.  It depends only on the $B$ decay angle,
defined as the angle $\theta$ between the two leptons in the $B$ rest frame, through:
\beq
  \label{eq:mll}
  m_{ll}^2=\frac{1}{2} x (1-y)(1-z)(1-\cos\theta)m_D^2.
\eeq
We define therefore the rescaled dilepton invariant mass
\beq
  \label{eq:mllhat}
  \mhat_{ll}=m_{ll}/(m_{ll})_{\mathrm{max}}=\sin(\theta/2).
\eeq

\begin{figure}[!htbp]
  \centering
  \includegraphics[width=0.47\textwidth]{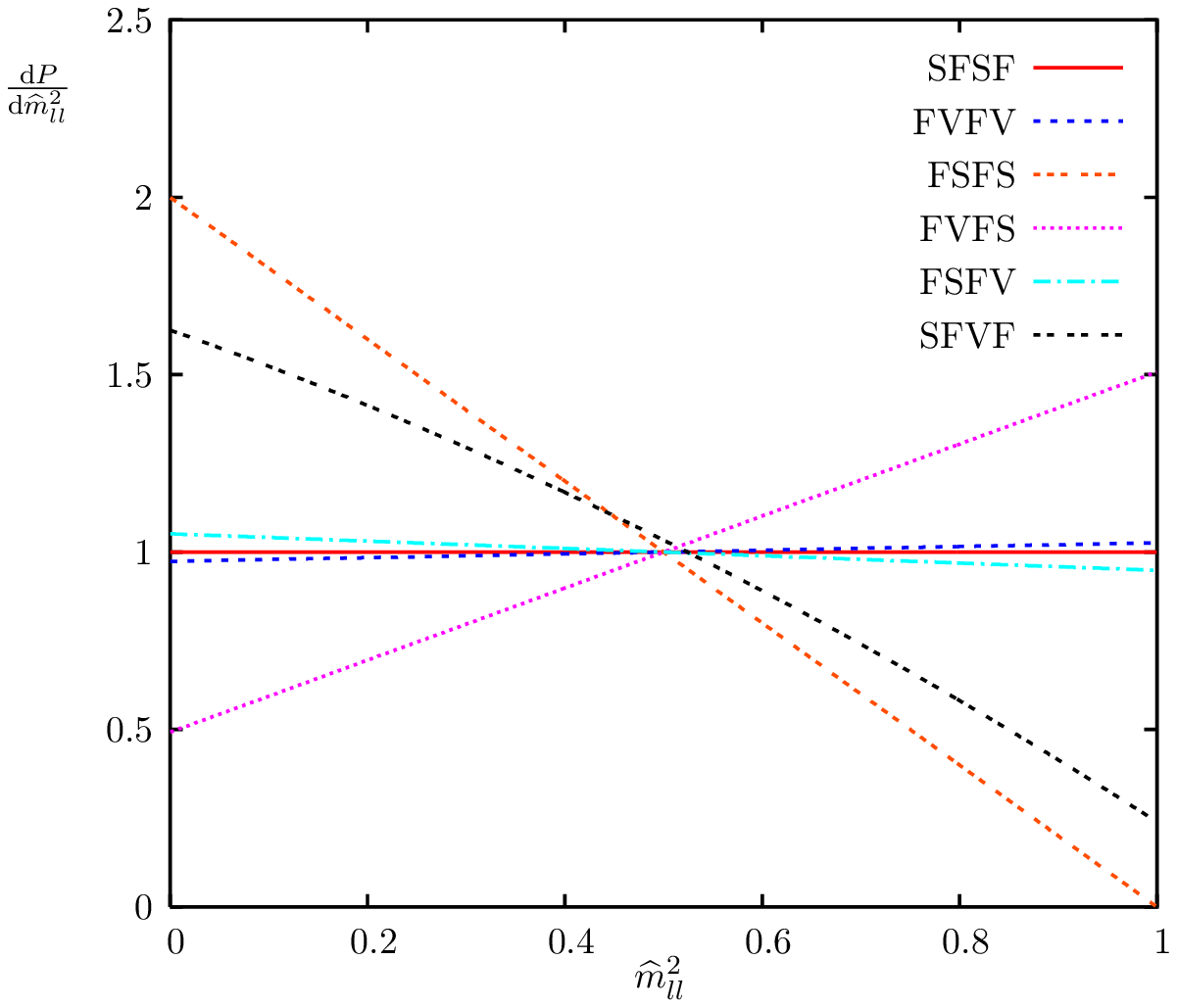}
  \includegraphics[width=0.47\textwidth]{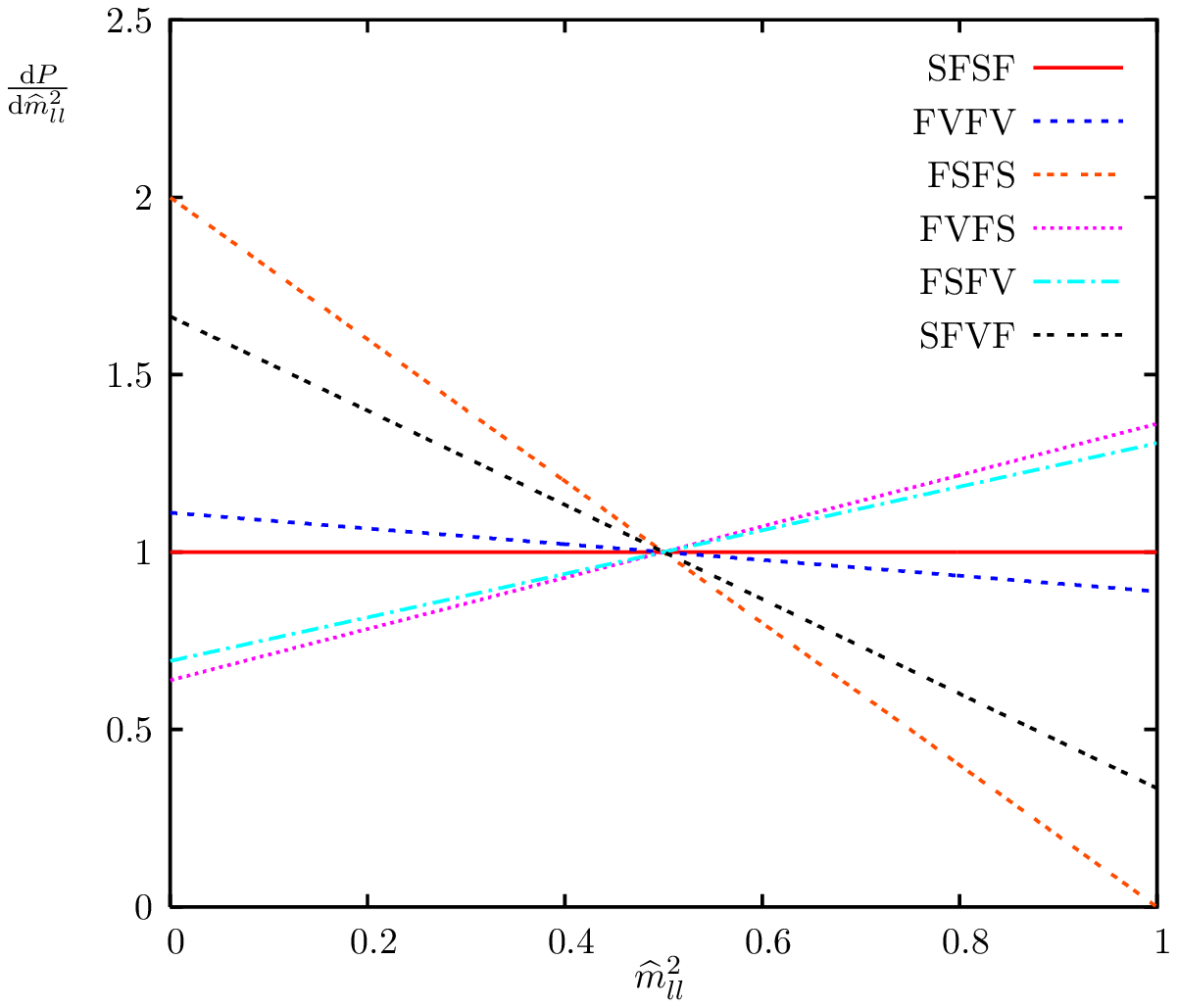}
  \setlength{\unitlength}{1cm}
  \begin{picture}(15,0.5)
    \put(4,0){(a)}
    \put(11.5,0){(b)}
  \end{picture}
  \caption{Dilepton mass distributions for (a) mass spectrum I and (b) mass spectrum II.} 
  \label{fig:ll}
\end{figure}

Figure \ref{fig:ll} shows the dilepton mass distribution, $\ud P/\ud \mhat_{ll}^2$, as a function
of $\mhat_{ll}^2$ for the 6 decay chains under consideration for mass spectra I and II.  The
analytical equations for the functions are in appendix \ref{sec:llPDFS}.  Figures
\ref{fig:ll} -- \ref{fig:qllmean} are plotted as functions of $\mhat^2$, as opposed to
functions of $\mhat$ as was done in \cite{Smillie:2005ar}, as this makes it easier to see
the functional dependence.

We see from figure~\ref{fig:ll} that, as found in \cite{Smillie:2005ar}, the SFSF (SUSY)
and FVFV (UED) decay chains would be very hard to distinguish on the basis of the dilepton
distribution.  On the other hand the FSFS and FVFS cases, where one or both of the UED
vector particles is replaced by a scalar, are characteristically different, as is the
chain in which one SUSY scalar is replaced by a vector (SFVF). Here and in the discussion
of subsequent plots, we shall quantify these initial qualitative observations in
section~\ref{sec:model-discrimination}.

\subsection{Quark and near lepton mass distributions}
\label{sec:quark-near-lepton}
The quark and near lepton distribution is not experimentally observable as the near and
far leptons cannot be distinguished.  We can however measure jet $l^{\pm}$ mass
distributions, as pointed out in \cite{Barr:2004ze}.  In order to compute these, we must
first calculate the near and far distributions.  These are then combined in section
\ref{sec:observ-quark-lept}.

The quark and near lepton invariant mass, $m_{ql}^{\rm near}$, is given in terms of the
angle between the two particles, $\theta^*$, in the rest frame of particle $C$:
\beq
  \label{eq:mqln}
  (m_{ql}^{\rm near})^2=\frac{1}{2} (1-x)(1-y)(1-\cos\theta^*)m_D^2.
\eeq
We then define the rescaled invariant mass
\beq
    \mhat_{ql}^{\rm near}=m_{ql}^{\rm near}/(m_{ql}^{\rm near})_{\rm max}=\sin(\theta^*/2).
    \label{eq:mqlnhat}
\eeq

Figure \ref{fig:qln1} shows the quark and near lepton mass distribution, $\ud P/\ud
(\mhat_{ql}^{\rm near})^2$, in process 1 as a function of $(\mhat_{ql}^{\rm near})^2$ for mass
spectra I and II.  Figure \ref{fig:qln2} shows the same thing for process 2.  These are
reflections of the distributions for process 1 about the point $(\mhat_{ql}^{\rm near})^2=1/2$.  The
analytical equations for the functions are in appendix \ref{sec:qlnPDFS}.

\FIGURE{
  \centering
  \includegraphics[width=0.47\textwidth]{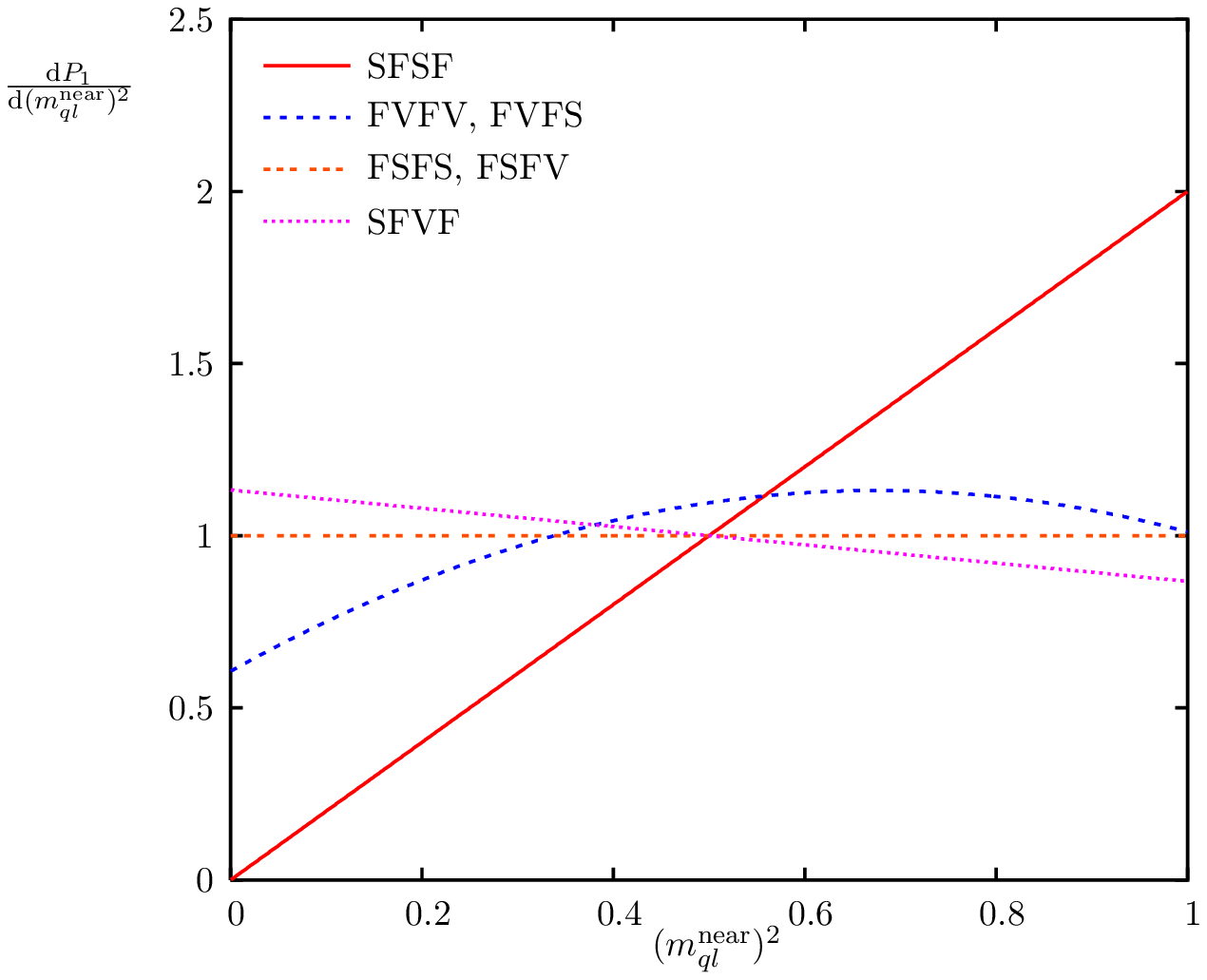}
  \includegraphics[width=0.47\textwidth]{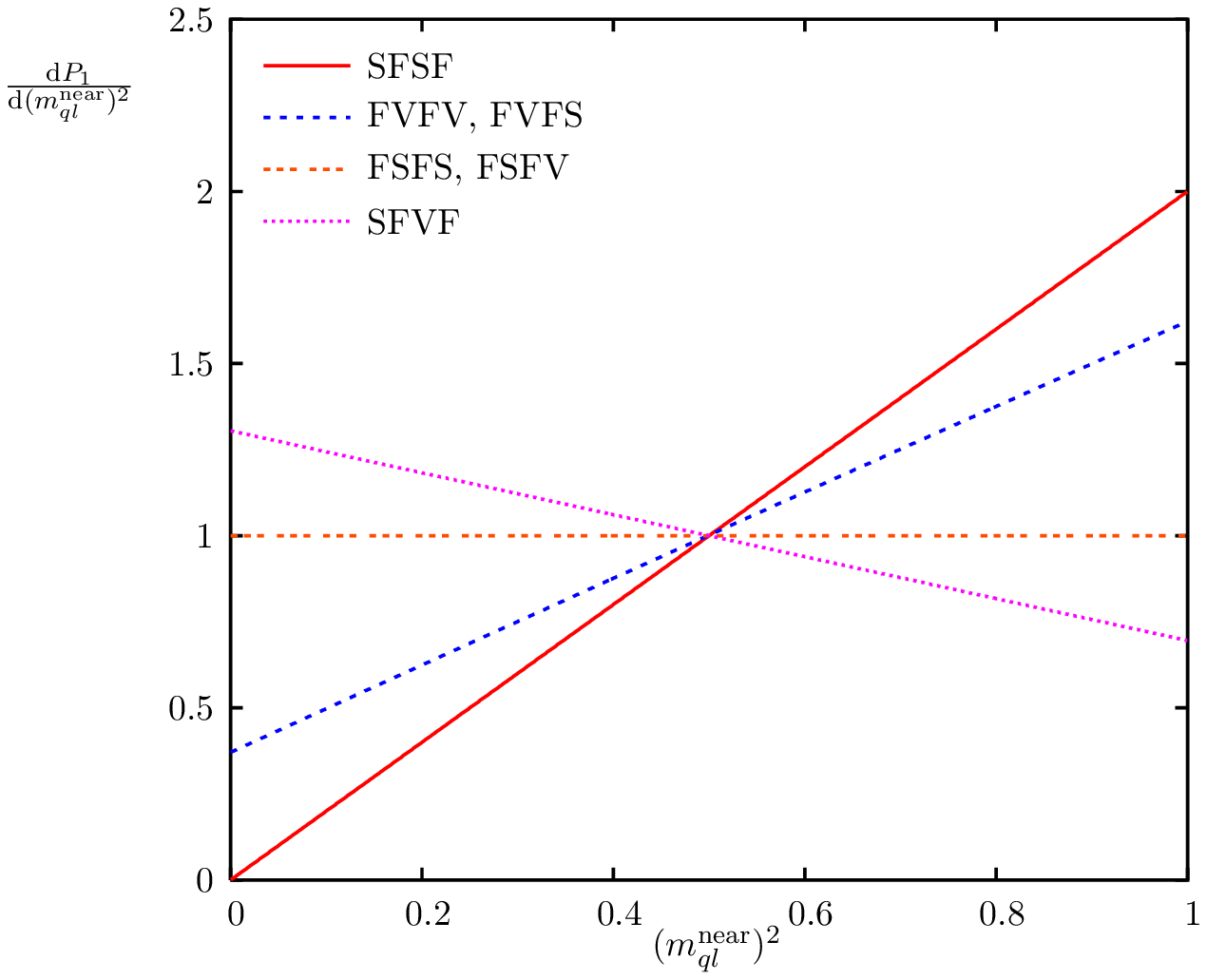}
    \setlength{\unitlength}{1cm}
  \begin{picture}(15,0.5)
    \put(4,0){(a)}
    \put(11.5,0){(b)}
  \end{picture}
  \caption{Quark and near lepton mass distributions for process 1 with (a) mass spectrum I
    and (b) mass spectrum II.}
  \label{fig:qln1}
}

\FIGURE{
  \centering
  \includegraphics[width=0.47\textwidth]{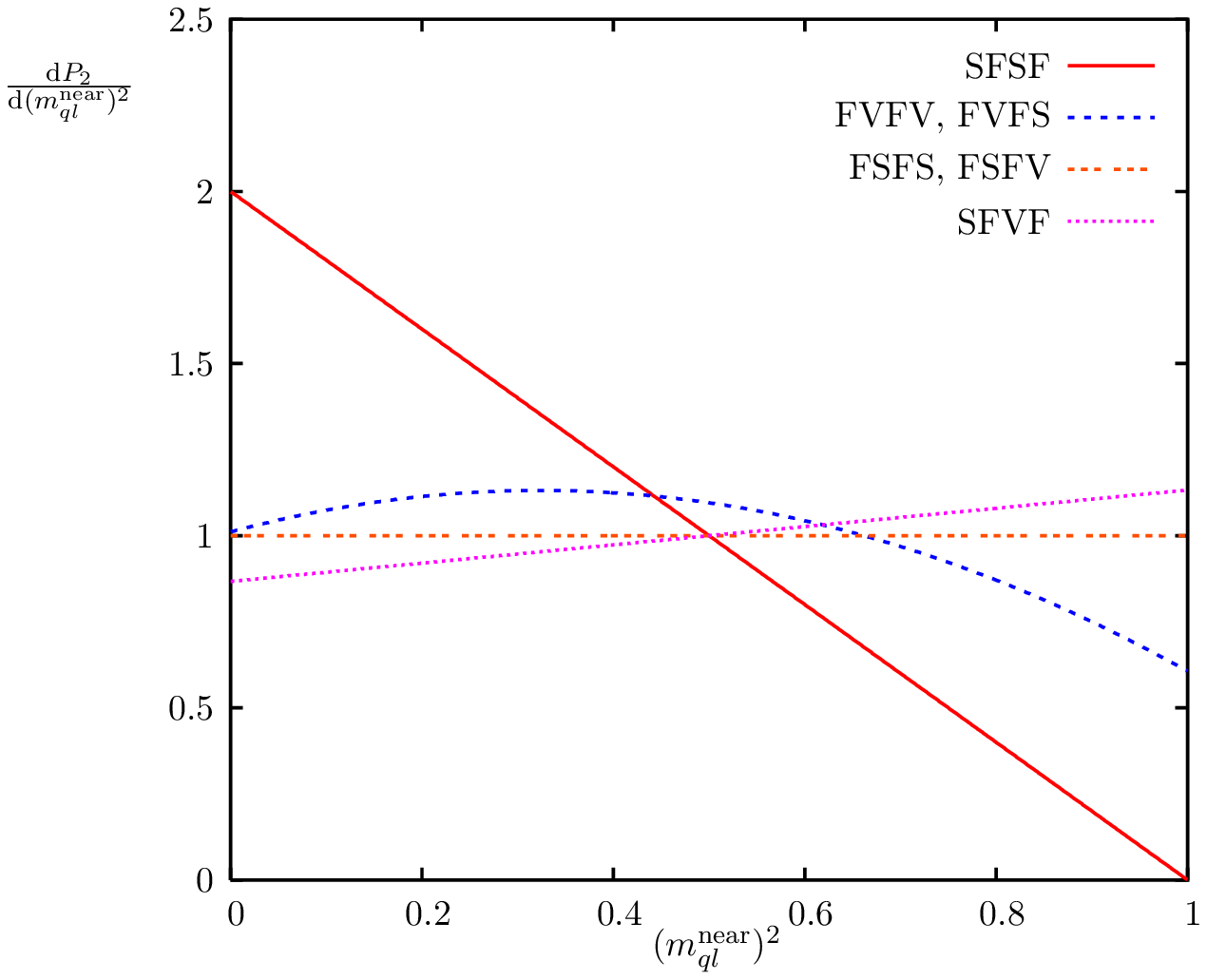}
  \includegraphics[width=0.47\textwidth]{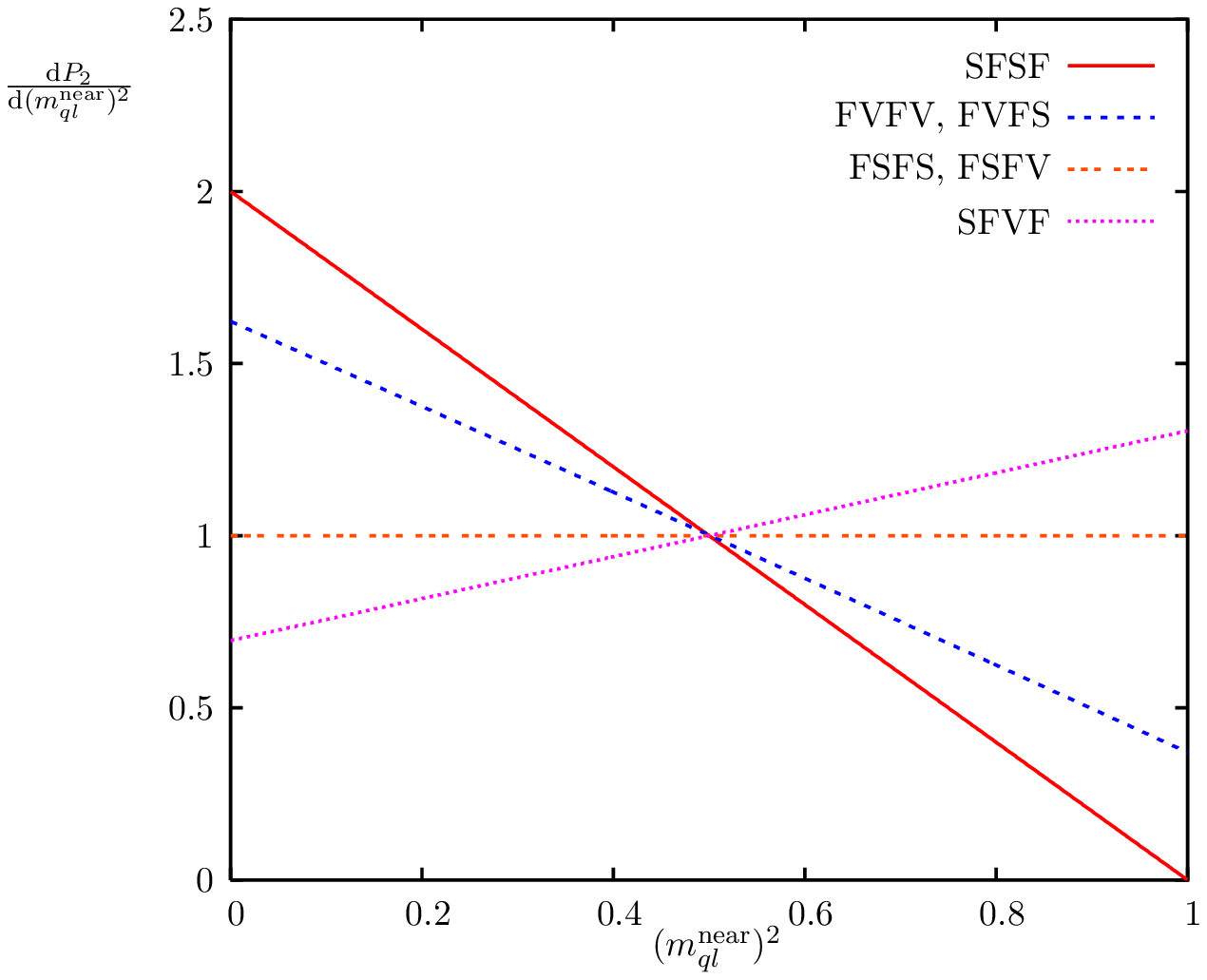}
  \setlength{\unitlength}{1cm}
  \begin{picture}(15,0.5)
    \put(4,0){(a)}
    \put(11.5,0){(b)}
  \end{picture}
  \caption{Quark and near lepton mass distributions for process 2 with (a) mass spectrum I
    and (b) mass spectrum II.}
  \label{fig:qln2}
}

Note that the distributions for FVFV and FVFS and for FSFS and FSFV are the same as the
differences in the chains only occur at the last vertex, which has no effect on the
quark and near lepton distribution.  Furthermore the distribution for FSFS and FSFV is
flat since the quark and near lepton are connected by a scalar in these cases.
The prospects for distinguishing the remaining cases look quite favourable.
However, as remarked above, we must first take account of the contribution
of the far lepton.

\subsection{Quark and far lepton mass distributions}
\label{sec:quark-far-lepton}

The quark and far lepton mass is a more complicated expression than the other two.  It is
a function of all the decay angles in the chain:
\begin{eqnarray}
  \label{eq:mqlf}
  (m_{ql}^{\rm far})^2&=&\frac{1}{4}(1-x)(1-z)\left[
    (1+y)(1-\cos\theta^*\cos\theta) \right. \nonumber \\ && \quad \left.
    +(1-y)(\cos\theta^*-\cos\theta)-2\sqrt{y} \sin\theta^*\sin\theta\cos\phi \right] m_D^2,
\end{eqnarray}
where $\theta^*$ and $\theta$ are as before and $\phi$ is the angle between the $ql^{\rm
  near}$ and dilepton planes, in the rest frame of $B$.  The maximum value is at
$\theta^*=0$ and $\theta=\pi$ giving 
\beq
(m_{ql}^{\rm far})_{\rm max}^2=(1-x)(1-z)m_D^2,
\eeq
so we define the rescaled mass to be
\begin{eqnarray}
  \label{eq:mhatqlf}
  \mhat_{ql}^{\rm far} \equiv (m_{ql}^{\rm far})/(m_{ql}^{\rm far})_{\rm max} &=& \frac{1}{2} \left[
    (1+y)(1-\cos\theta^*\cos\theta) + \right. \nonumber \\ && \quad
  \left. (1-y)(\cos\theta^*-\cos\theta)-2\sqrt{y} \sin\theta^*\sin\theta\cos\phi \right]^{\frac{1}{2}}.
\end{eqnarray}

Figure \ref{fig:qlf1} shows the quark and far lepton mass distribution, $\ud P/\ud
(\mhat_{ql}^{\rm far})^2$, in process 1 as a function of $(\mhat_{ql}^{\rm far})^2$ for mass
spectra I and II.  Figure \ref{fig:qlf2} shows the same thing for process 2.  The
analytical equations for the functions are in appendix \ref{sec:qlfPDFS}.

\FIGURE{
  \centering
  \includegraphics[width=0.47\textwidth]{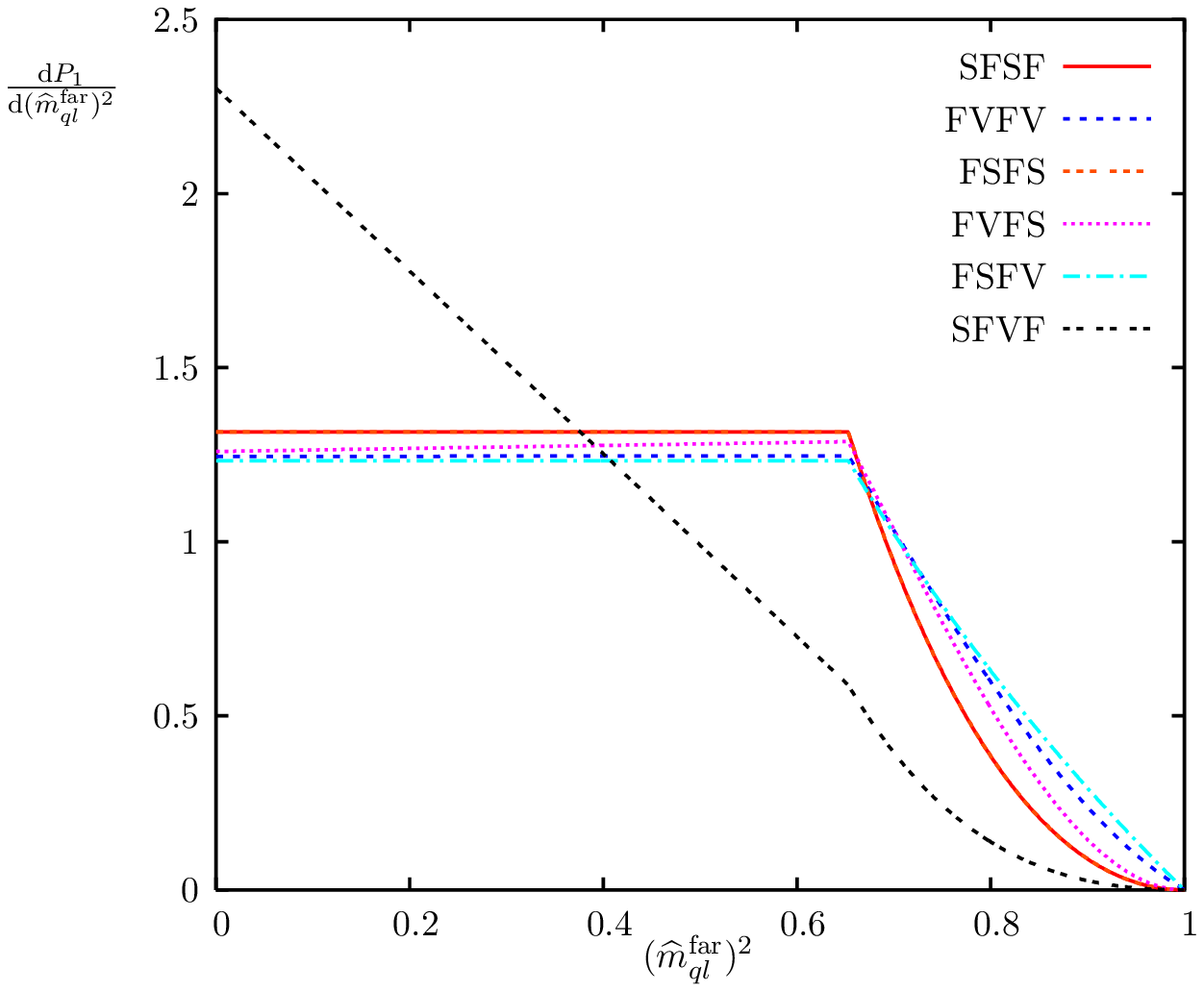}
  \includegraphics[width=0.47\textwidth]{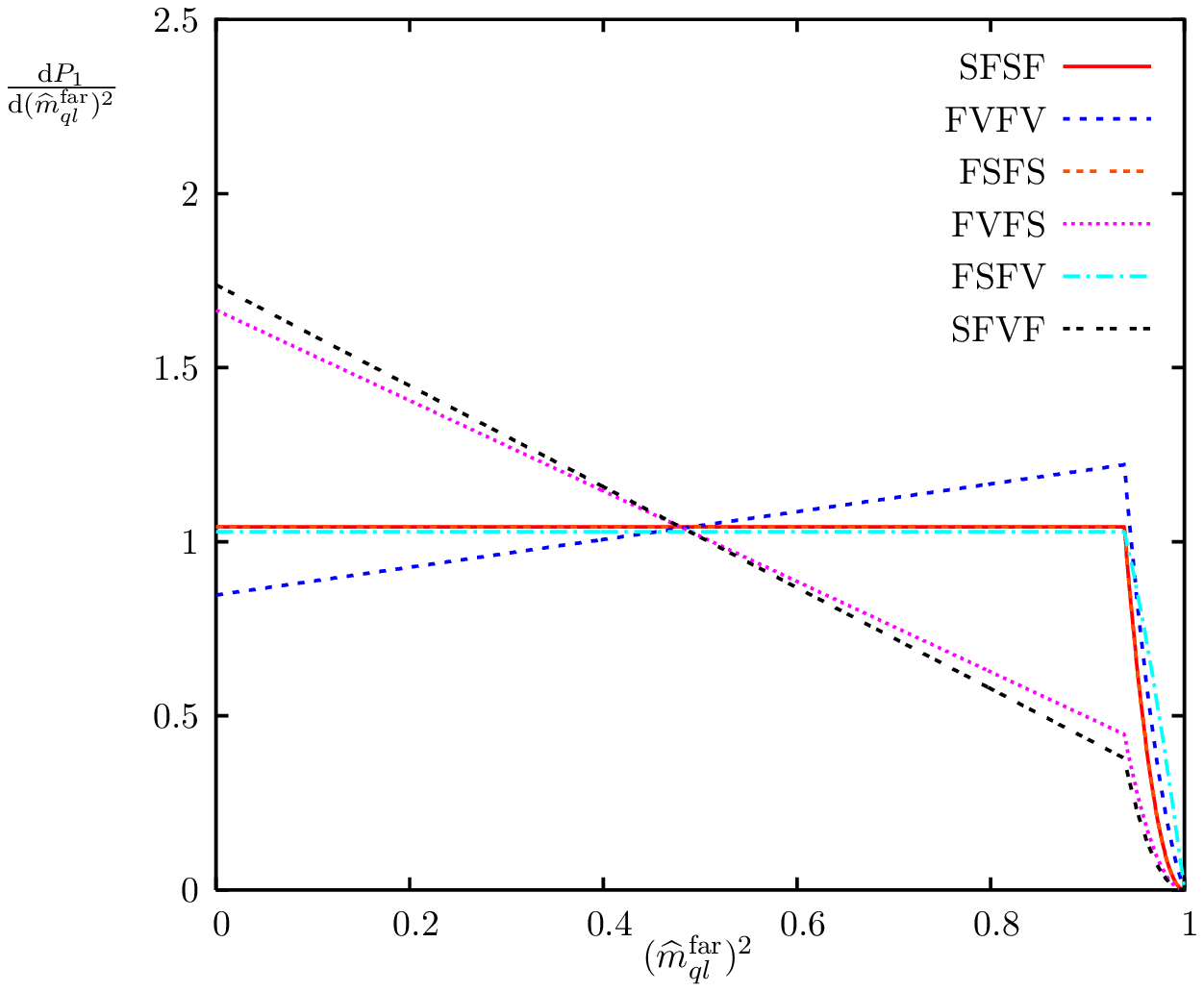}
  \setlength{\unitlength}{1cm}
  \begin{picture}(15,0.5)
    \put(4,0){(a)}
    \put(11.5,0){(b)}
  \end{picture}
  \caption{Quark and far lepton mass distributions for process 1 with (a) mass spectrum I
    and (b) mass spectrum II.}
  \label{fig:qlf1}
}

\FIGURE{
  \centering
  \includegraphics[width=0.47\textwidth]{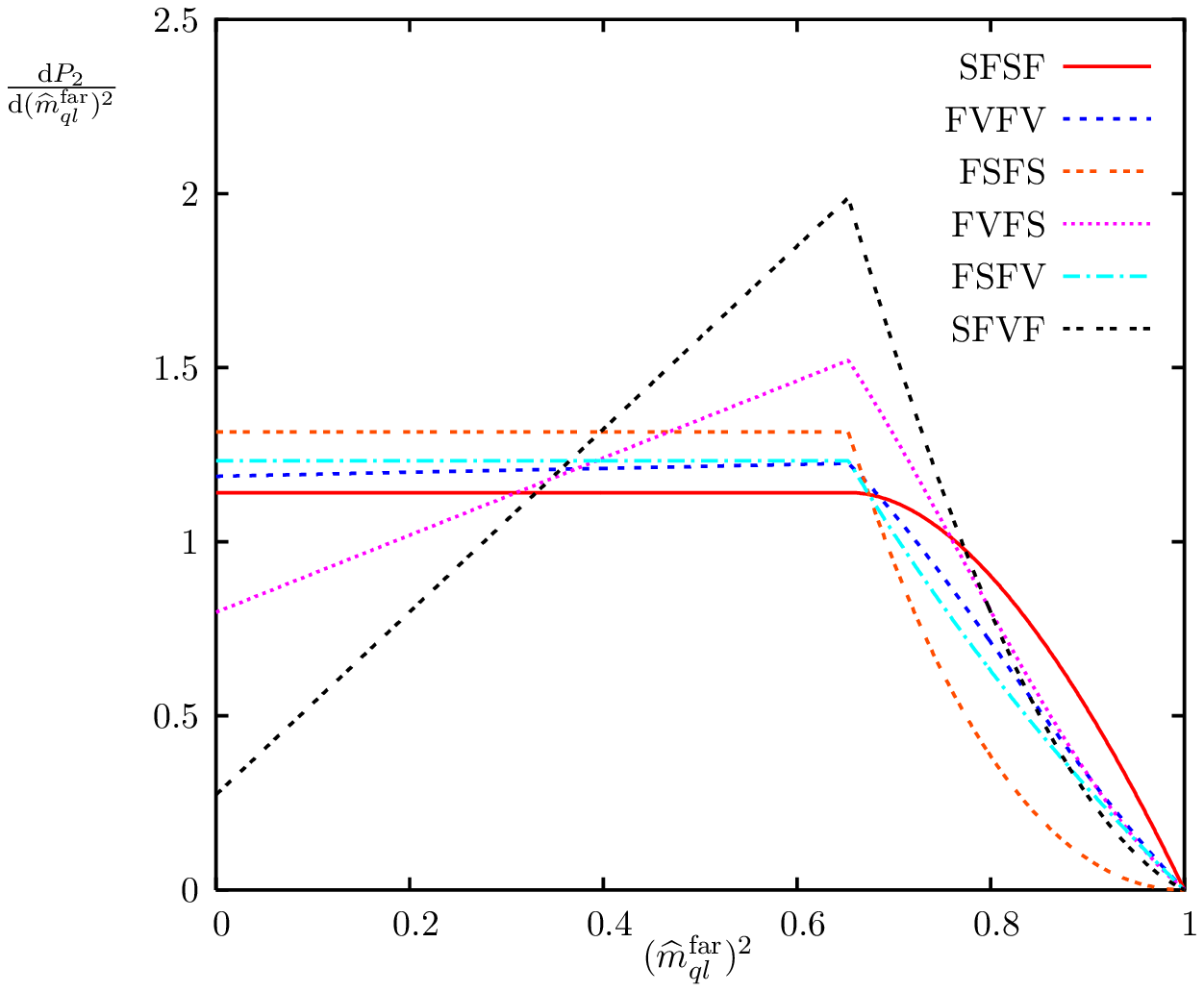}
  \includegraphics[width=0.47\textwidth]{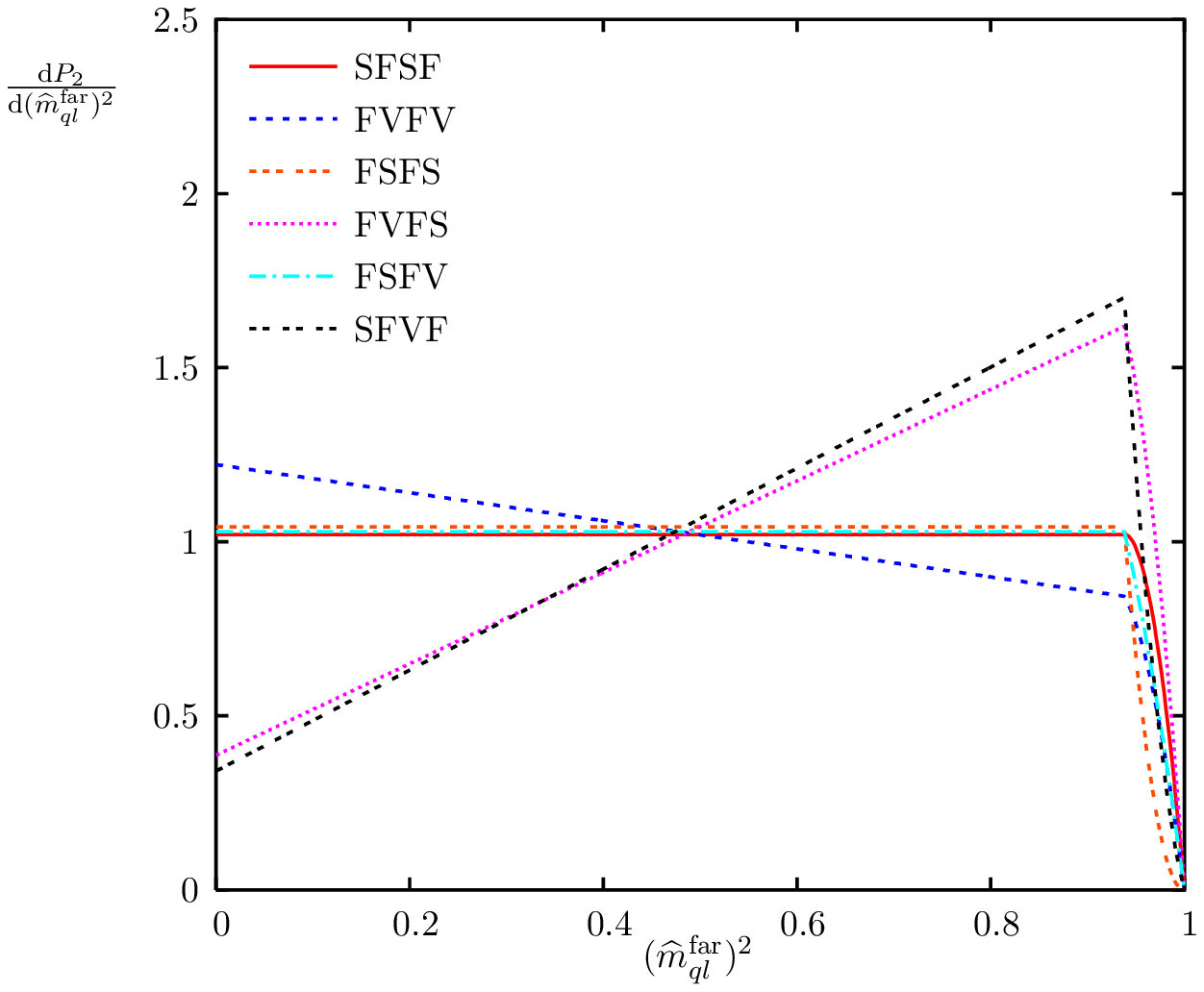}
  \setlength{\unitlength}{1cm}
  \begin{picture}(15,0.5)
    \put(4,0){(a)}
    \put(11.5,0){(b)}
  \end{picture}
  \caption{Quark and far lepton mass distributions for process 2 with (a) mass spectrum I
    and (b) mass spectrum II.}
  \label{fig:qlf2}
}

It can be seen from figures \ref{fig:qlf1} and \ref{fig:qlf2}, and from the formulae in
appendix~\ref{sec:qlfPDFS}, that in many cases the dependence on the quark plus far lepton
mass is absent or weak throughout the region $0\le (\mhat_{ql}^{\rm far})^2\le y$. The
main exception is SFVF, where the far lepton is connected to the chain via a vector
particle.  At higher values of $\mhat_{ql}^{\rm far}$, the discrimination is marginally
better, which is naturally more useful for spectra with smaller values of $y$ (mass
spectrum I).  It is clear, however, that the contribution from the far lepton will degrade
the power of the quark plus lepton distribution to distinguish between spin assignments.

\subsection{Observable quark-lepton mass distribution}
\label{sec:observ-quark-lept}
We now consider the jet + lepton combinations, $jl^{\pm}$, as first used in
\cite{Barr:2004ze}.  As in \cite{Smillie:2005ar}, we assume that the jet and lepton are
indeed decay products from process 1 or process 2.  Then the $jl^{\pm}$ mass distribution
for a given lepton charge receives near- and far-lepton contributions from both the
corresponding process and the charge conjugate of the other process.  In other words, we
have
\beq\label{eq:mjl+}
\frac{\ud P}{\ud m_{jl^+}}
= \frac 12 \left[ f_q\left(
\frac{\ud P_2}{\ud m_{ql}^{\rm near}}+ 
\frac{\ud P_1}{\ud m_{ql}^{\rm far}}\right)
+ f_{\bar q}\left(
\frac{\ud P_1}{\ud m_{ql}^{\rm near}}+ 
\frac{\ud P_2}{\ud m_{ql}^{\rm far}}\right)\right]
\eeq
where $f_q$ and $f_{\bar q}$ are the quark and antiquark fractions
in the selected event sample. The factor of one-half enters because
$P_{1,2}$ are both normalized to unity.  Similarly
\beq\label{eq:mjl-}
\frac{\ud P}{\ud m_{jl^-}}
= \frac 12 \left[ f_q\left(
\frac{\ud P_1}{\ud m_{ql}^{\rm near}}+ 
\frac{\ud P_2}{\ud m_{ql}^{\rm far}}\right)
+ f_{\bar q}\left(
\frac{\ud P_2}{\ud m_{ql}^{\rm near}}+ 
\frac{\ud P_1}{\ud m_{ql}^{\rm far}}\right)\right]\;.
\eeq
This has assumed that both of the leptons are left-handed --
in the case of right-handed leptons, the expressions for $jl^+$ and $jl^-$ are
interchanged.

In \cite{Smillie:2005ar}, the \textsc{Herwig} event generator
\cite{Corcella:2000bw,Corcella:2002jc} was used to numerically calculate $f_q$ in the
cases of SUSY and UED for two different mass spectra.  Despite the differences in the
models, both gave $f_q \approx 0.7$ for both mass spectra.  This is therefore the value
that we will take for all of our models.

Figures \ref{fig:jlp} and \ref{fig:jlm} show the resulting jet+lepton mass distributions
for mass spectra I and II, where we have normalised $\mhat_{jl}$ to the maximum observable
mass in each case.

\FIGURE{
  \centering
  \includegraphics[width=0.47\textwidth]{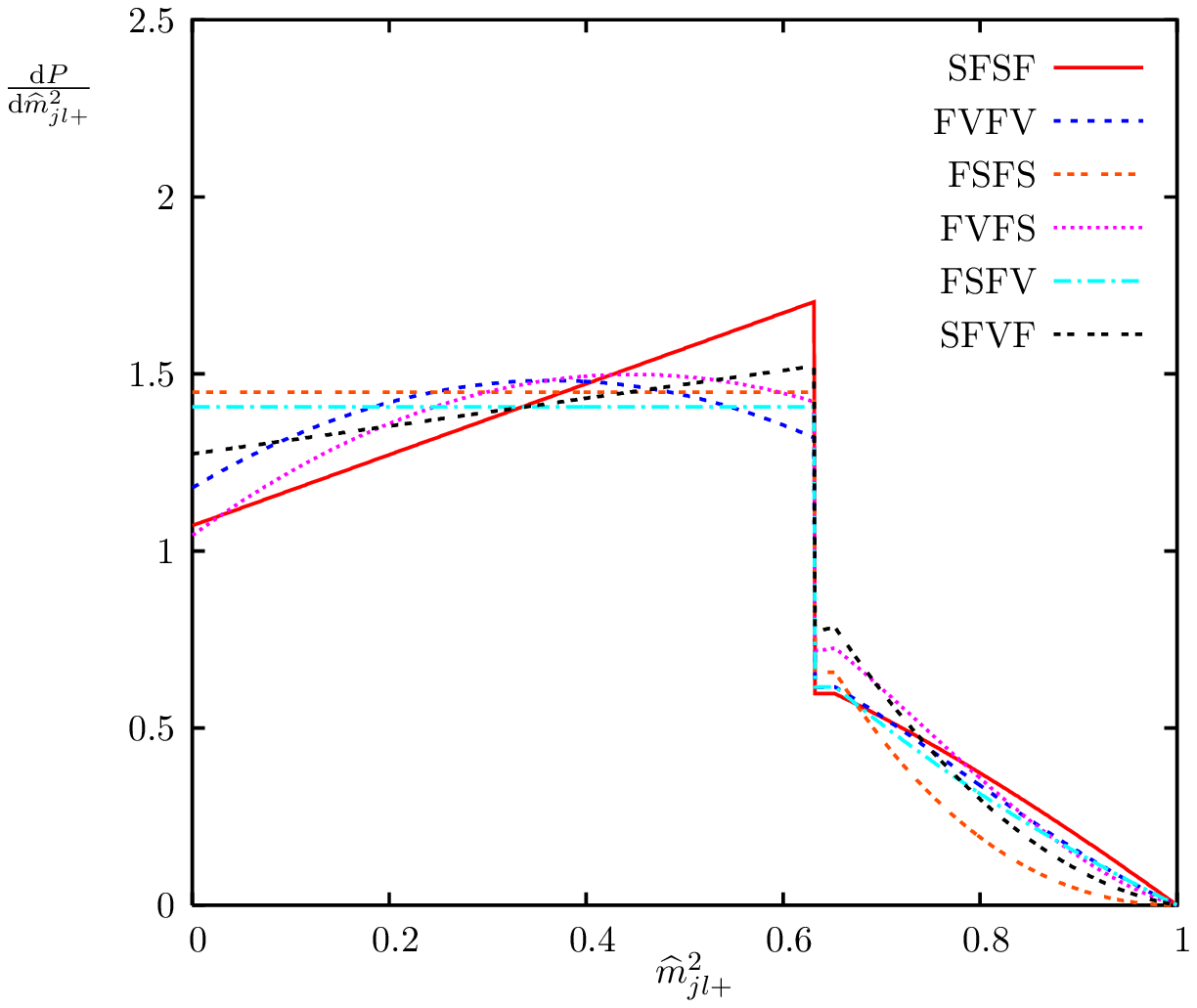}
  \includegraphics[width=0.47\textwidth]{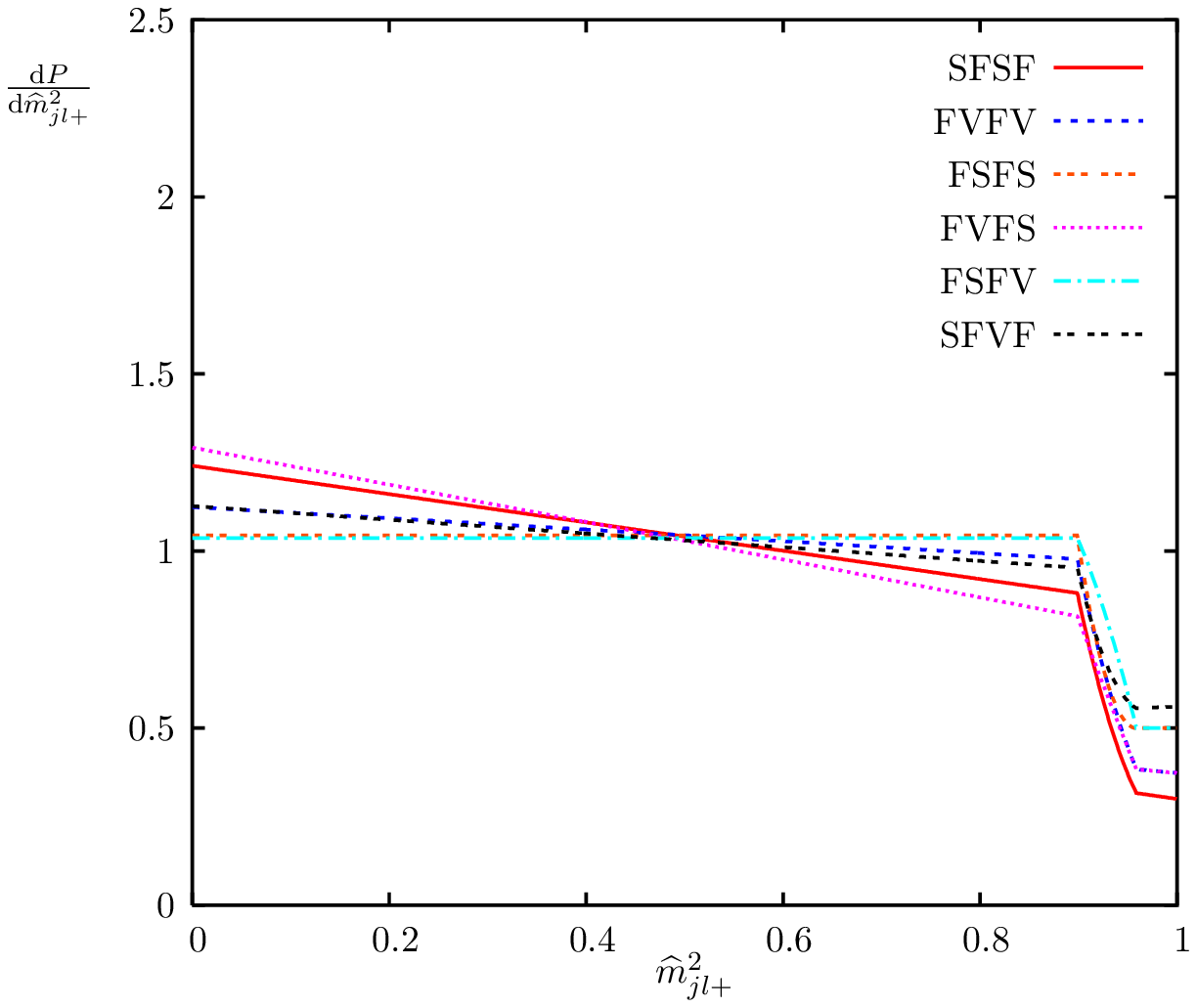}
  \setlength{\unitlength}{1cm}
  \begin{picture}(15,0.5)
    \put(4,0){(a)}
    \put(11.5,0){(b)}
  \end{picture}
  \caption{jet$+l^+$ mass distribution for (a) mass spectrum I and (b) mass spectrum II.}
  \label{fig:jlp}
}

\FIGURE{
  \centering
  \includegraphics[width=0.47\textwidth]{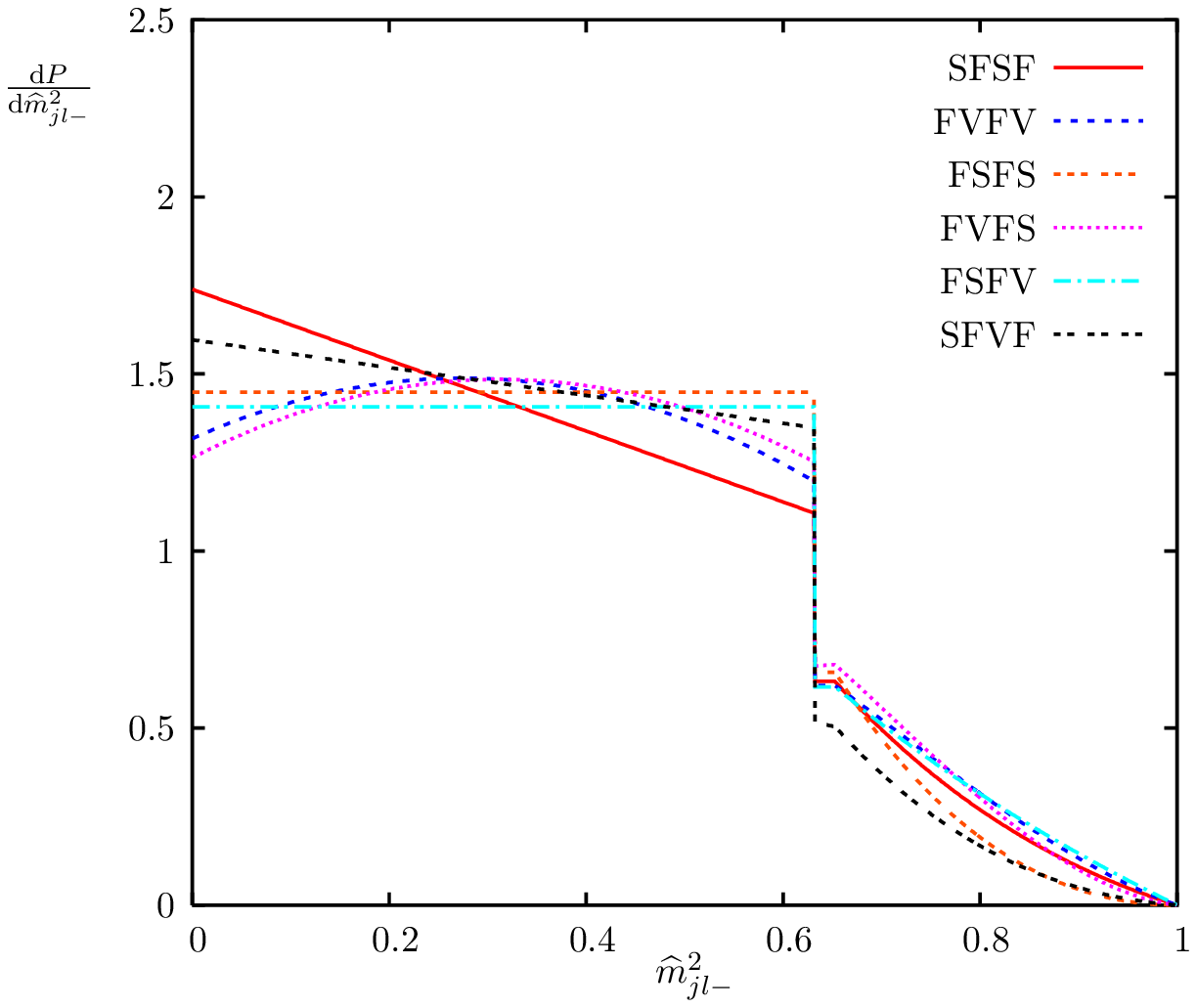}
  \includegraphics[width=0.47\textwidth]{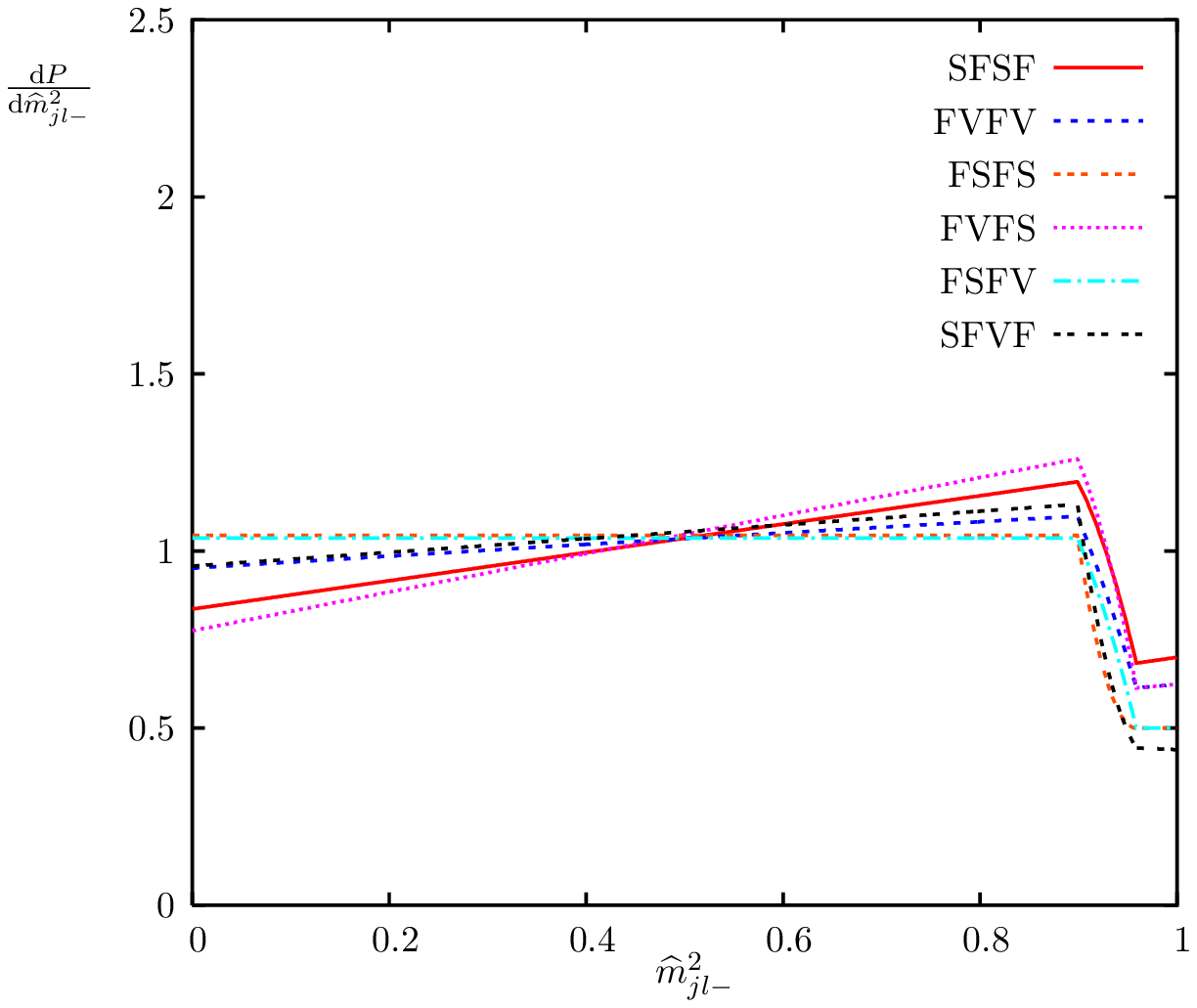}
  \setlength{\unitlength}{1cm}
  \begin{picture}(15,0.5)
    \put(4,0){(a)}
    \put(11.5,0){(b)}
  \end{picture}
  \caption{jet$+l^-$ mass distribution for (a) mass spectrum I and (b) mass spectrum II.}
  \label{fig:jlm}
}

The resulting charge asymmetry,
\beq
\label{eq:Apm}
A=\frac{\ud P/\ud m_{jl+}-\ud P/\ud m_{jl-}}{\ud P/\ud m_{jl+}+\ud P/\ud m_{jl-}},
\eeq
is plotted in figure \ref{fig:ApmFig}.  There is a reversal of sign
between the predictions for the two mass spectra, since, as discussed in
Section~\ref{sec:decay-chain}, the SUSY mass (I) plots assume
right-handed leptons, while the UED mass (II) plots  assume left.

\FIGURE{
  \centering
  \includegraphics[width=0.47\textwidth]{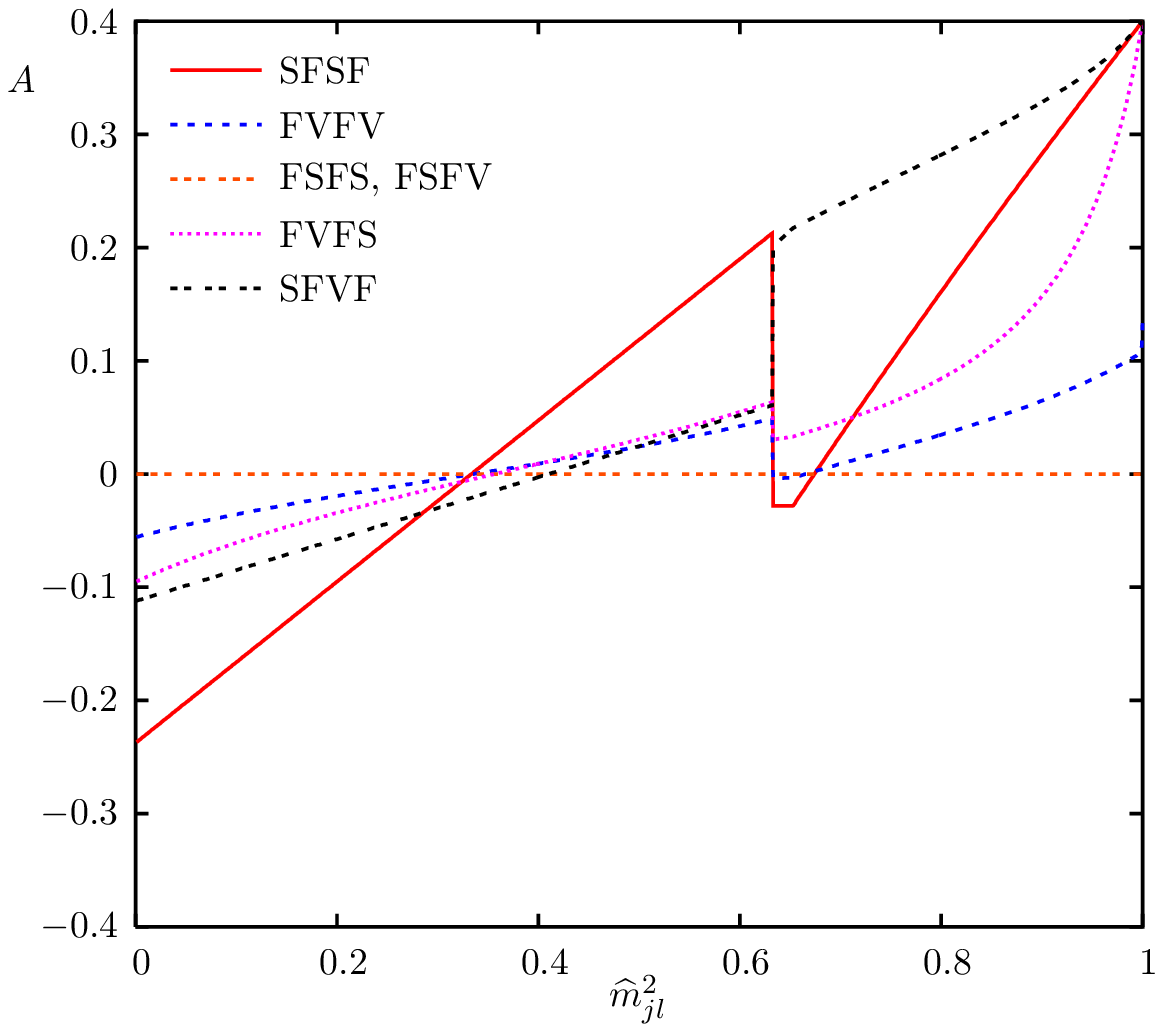}
  \includegraphics[width=0.47\textwidth]{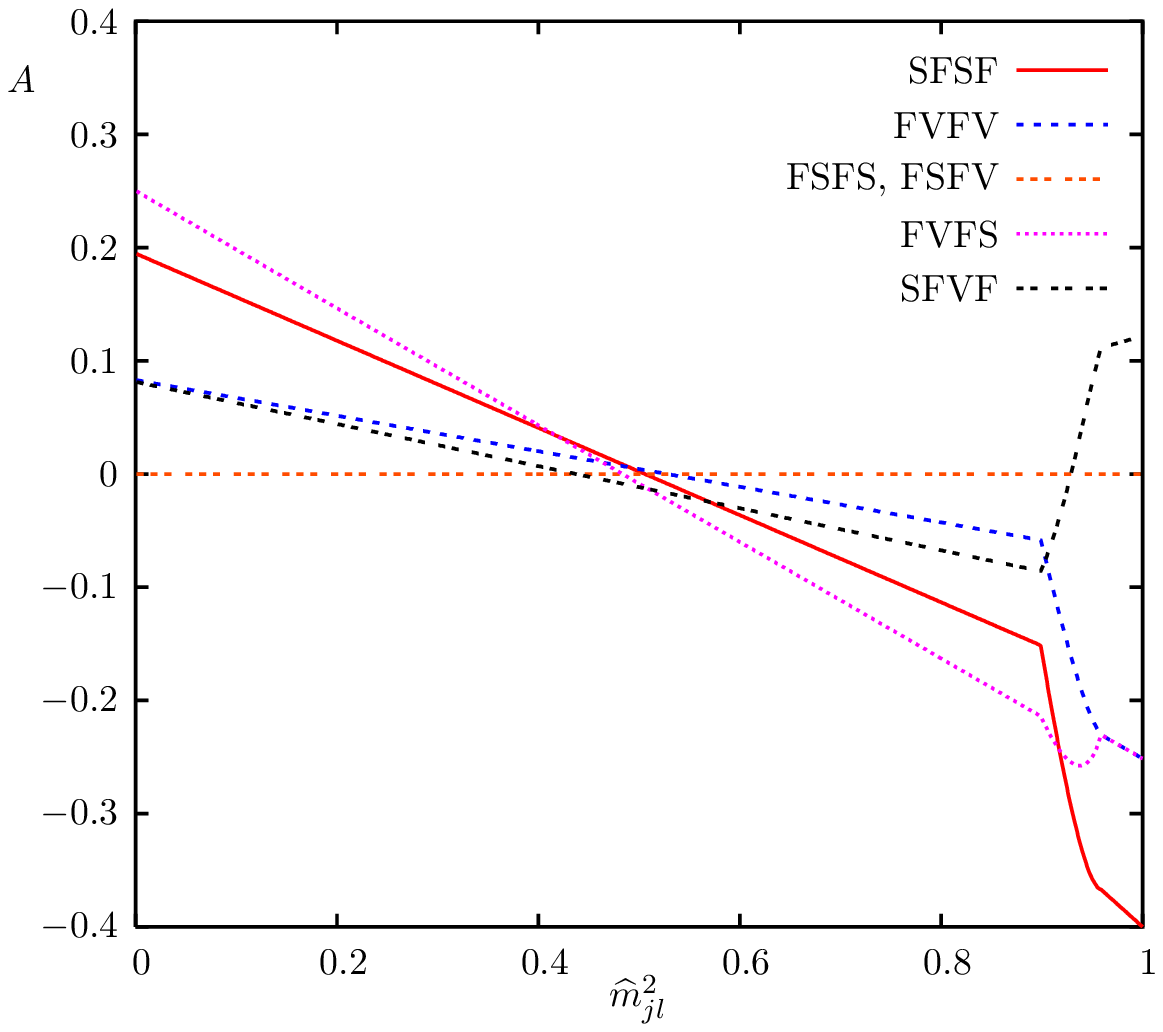}
  \setlength{\unitlength}{1cm}
  \begin{picture}(15,0.5)
    \put(4,0){(a)}
    \put(11.5,0){(b)}
  \end{picture}
  \caption{Asymmetry $A$ for (a) mass spectrum I and (b) mass spectrum II.}
  \label{fig:ApmFig}
}

It is clear from figures~\ref{fig:jlp} and \ref{fig:jlm} that high statistics would be
required to distinguish spin assignments on the basis of the jet plus lepton
distributions. This will be seen more quantitatively in section~\ref
{sec:model-discrimination}.  On the other hand, the asymmetries in figure~\ref{fig:ApmFig}
show striking characteristic differences that would provide convincing confirmation of a
model once sufficient events had been accumulated.

\subsection{Quark, near  and far lepton mass distribution}
\label{sec:quark-near-far}

The unscaled $ql^{\rm near}l^{\rm far}$ mass-squared is the sum of the three
pairwise unscaled masses-squared already considered:
\beq 
(m_{qll})^2=m_D^2\left[ x(1-y)(1-z)(\widehat{m}_{ll})^2 +
(1-x)(1-y)(\widehat{m}_{ql}^{\rm near})^2 + (1-x)(1-z) (\widehat{m}_{ql}^{\rm
  far})^2\right] .  
\label{eq:mqll}
\eeq 
From eqs. (\ref{eq:mllhat}), (\ref{eq:mqlnhat}) and (\ref{eq:mhatqlf}), $m_{qll}^2$
is a complicated function of the three angles $\theta,\theta^*$ and $\phi$.  The maximum
value of this depends upon the relative values of $x,y$ and $z$.  From
\cite{Allanach:2000kt,Lester:2001zx},
\beq 
(m_{qll}^{\rm max})^2= \left\{
  \begin{array}{c c}
    m_D^2(1-x)(1-yz) & \quad {\rm iff} \quad x<yz, \\
    m_D^2(1-xy)(1-z) & \quad {\rm iff} \quad z<xy, \\
    m_D^2(1-xz)(1-y) & \quad {\rm iff} \quad y<xz, \\
    m_D^2(1-\sqrt{xyz})^2 & \quad {\rm otherwise.}
  \end{array} \right.
\label{eq:mqllMAX}
\eeq
Both mass spectra I and II have $x<yz$; however, in order to remain independent of the
relative values of $x,y$ and $z$, we shall no longer
define $\widehat{m}_{qll}$ to lie between 0 and 1.  We shall instead simply define 
\beq
\widehat{m}_{qll}=m_{qll}/m_D,
\label{mqllhat}
\eeq
so that it is still dimensionless.  

Figure \ref{fig:qll1} shows the $qll$ mass distribution, $\ud P/\ud
(\widehat{m}_{qll})^2$, for process 1 for mass spectra I and II, while
Figure \ref{fig:qll2} shows the same thing for process 2.  The analytical
equations are discussed in appendix \ref{sec:qllPDFS}.

\FIGURE{
  \centering
  \includegraphics[width=0.47\textwidth]{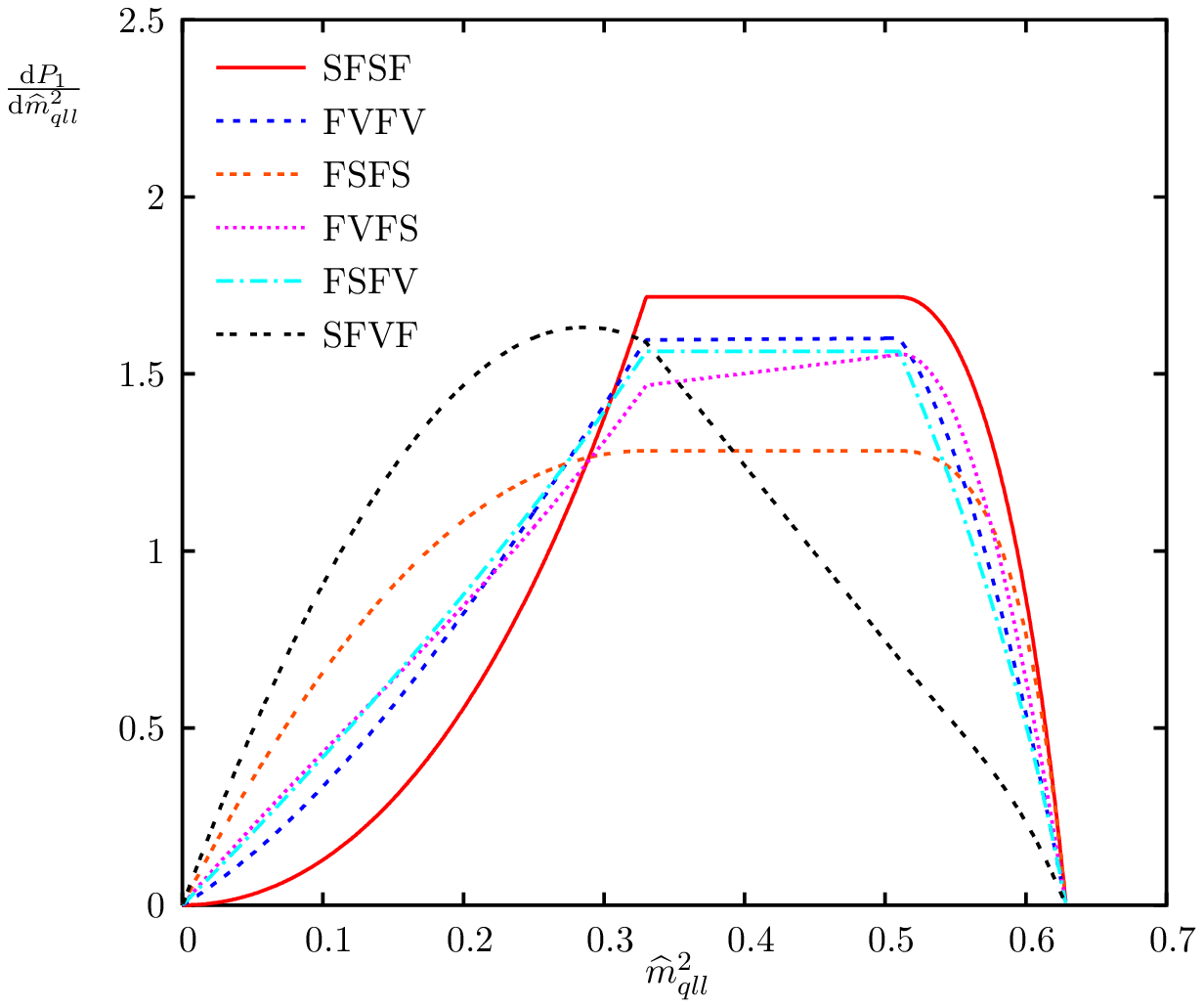}
  \includegraphics[width=0.47\textwidth]{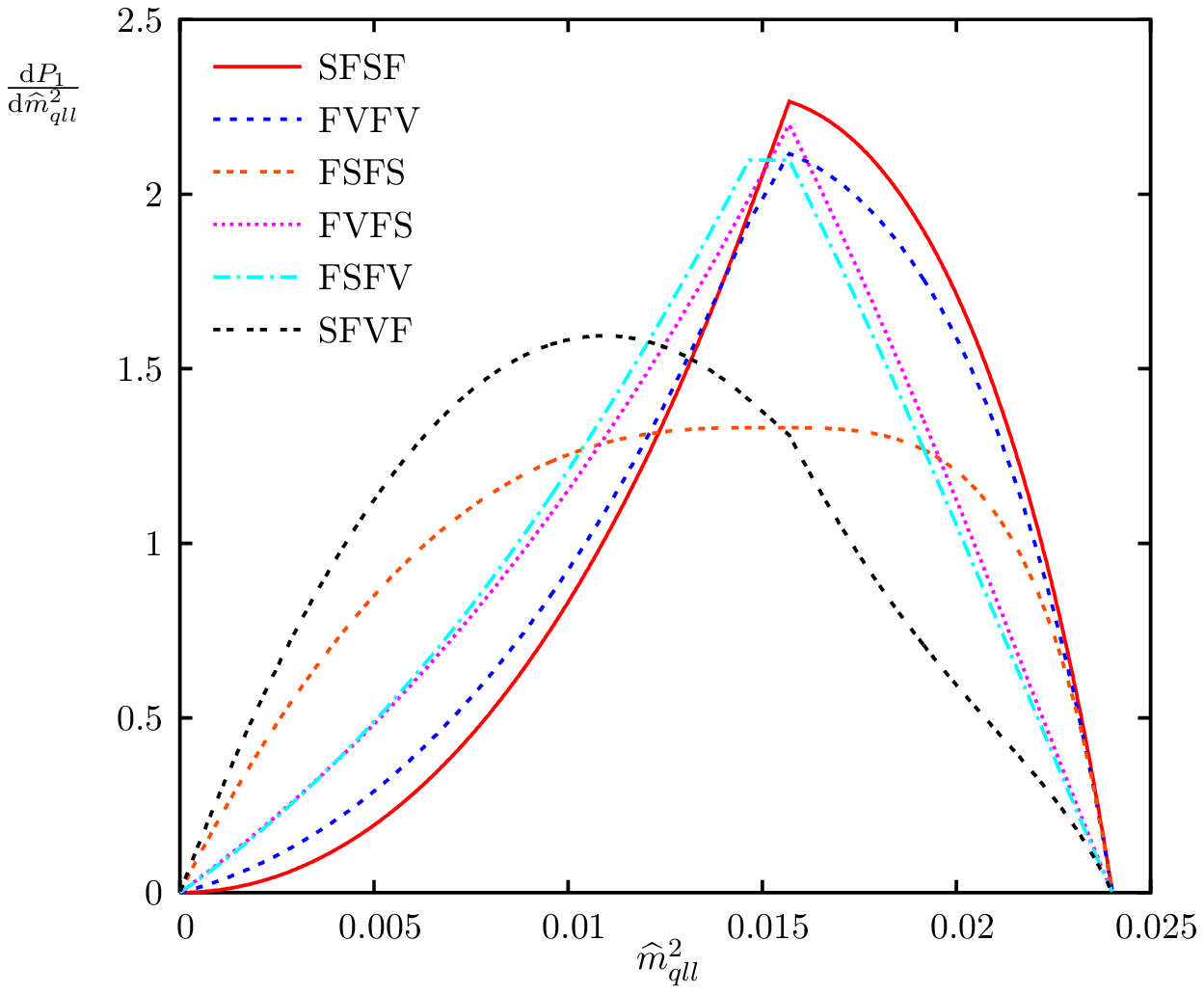}
  \setlength{\unitlength}{1cm}
  \begin{picture}(15,0.5)
    \put(4,0){(a)}
    \put(11.5,0){(b)}
  \end{picture}
  \caption{qll mass distributions for process 1 with (a) mass spectrum I and (b) mass
    spectrum II.}
  \label{fig:qll1}
}

\FIGURE{
  \centering
  \includegraphics[width=0.47\textwidth]{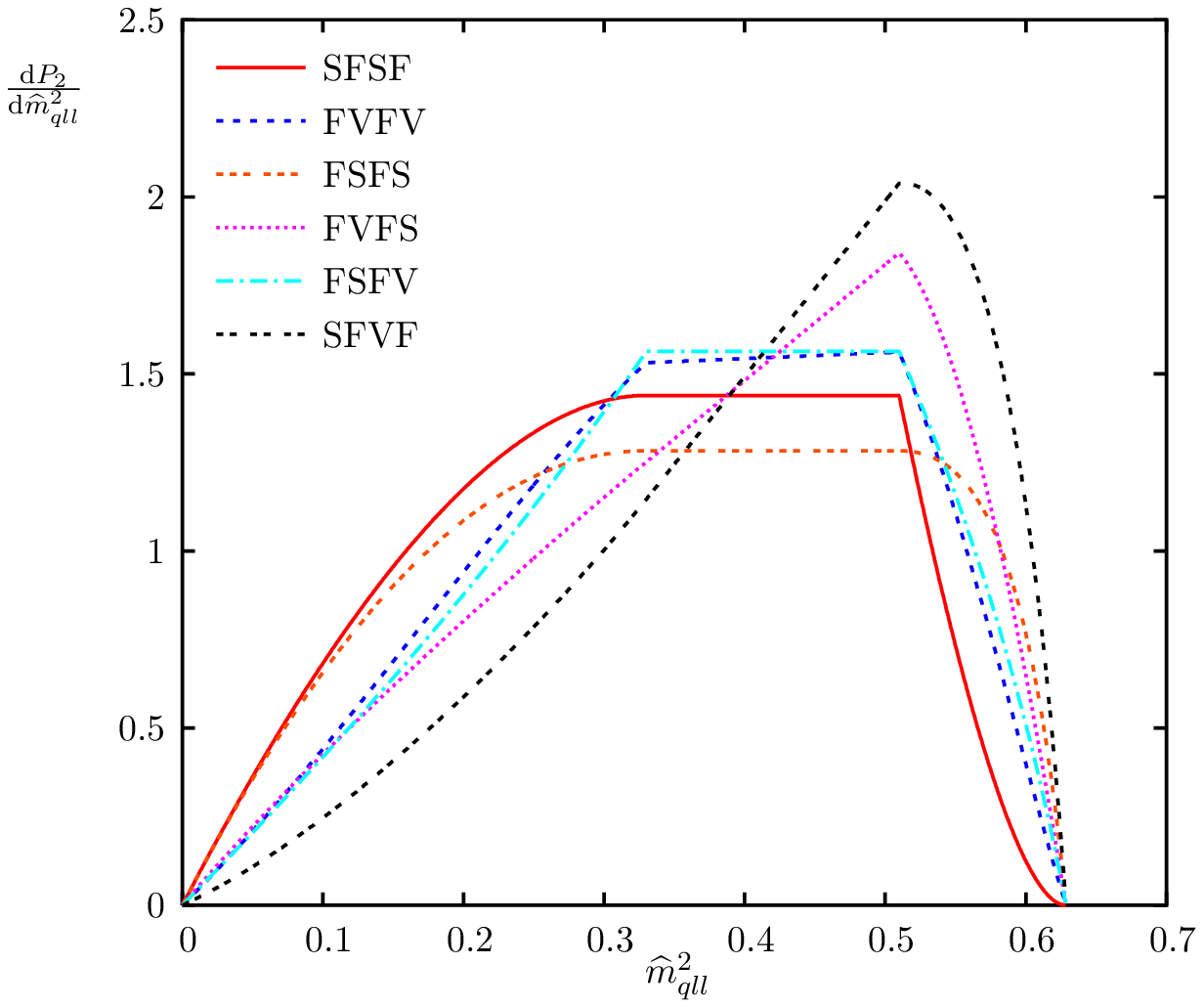}
  \includegraphics[width=0.47\textwidth]{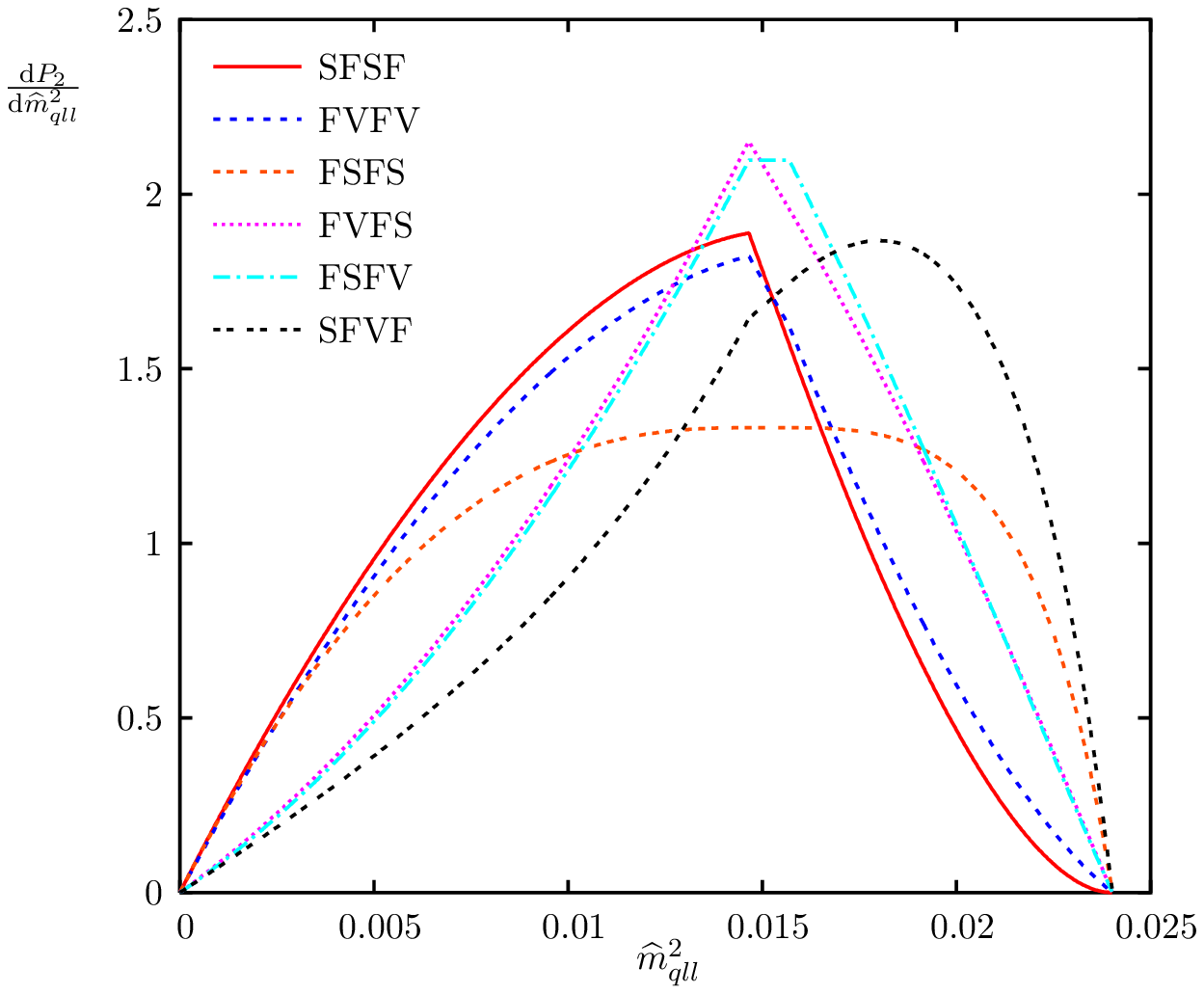}
  \setlength{\unitlength}{1cm}
  \begin{picture}(15,0.5)
    \put(4,0){(a)}
    \put(11.5,0){(b)}
  \end{picture}
  \caption{qll mass distributions for process 2 with (a) mass spectrum I and (b) mass
    spectrum 2.}
  \label{fig:qll2}
}

In an experimental situation we will be unable to distinguish between processes
1
and 2 and will instead see a jet+dilepton distribution that is an equal mixture of the
two. The asymmetry between
particle and antiparticle production at a $pp$ collider does not help here,
because the decays of the unstable particles in the chain into charge conjugate
modes must be equal.  Figure \ref{fig:qllmean} shows the combined $jll$ plots
for mass spectra I and II. We see that there is hope of distinguishing between
some cases in the intermediate region of invariant mass.

\FIGURE{
  \centering
  \includegraphics[width=0.47\textwidth]{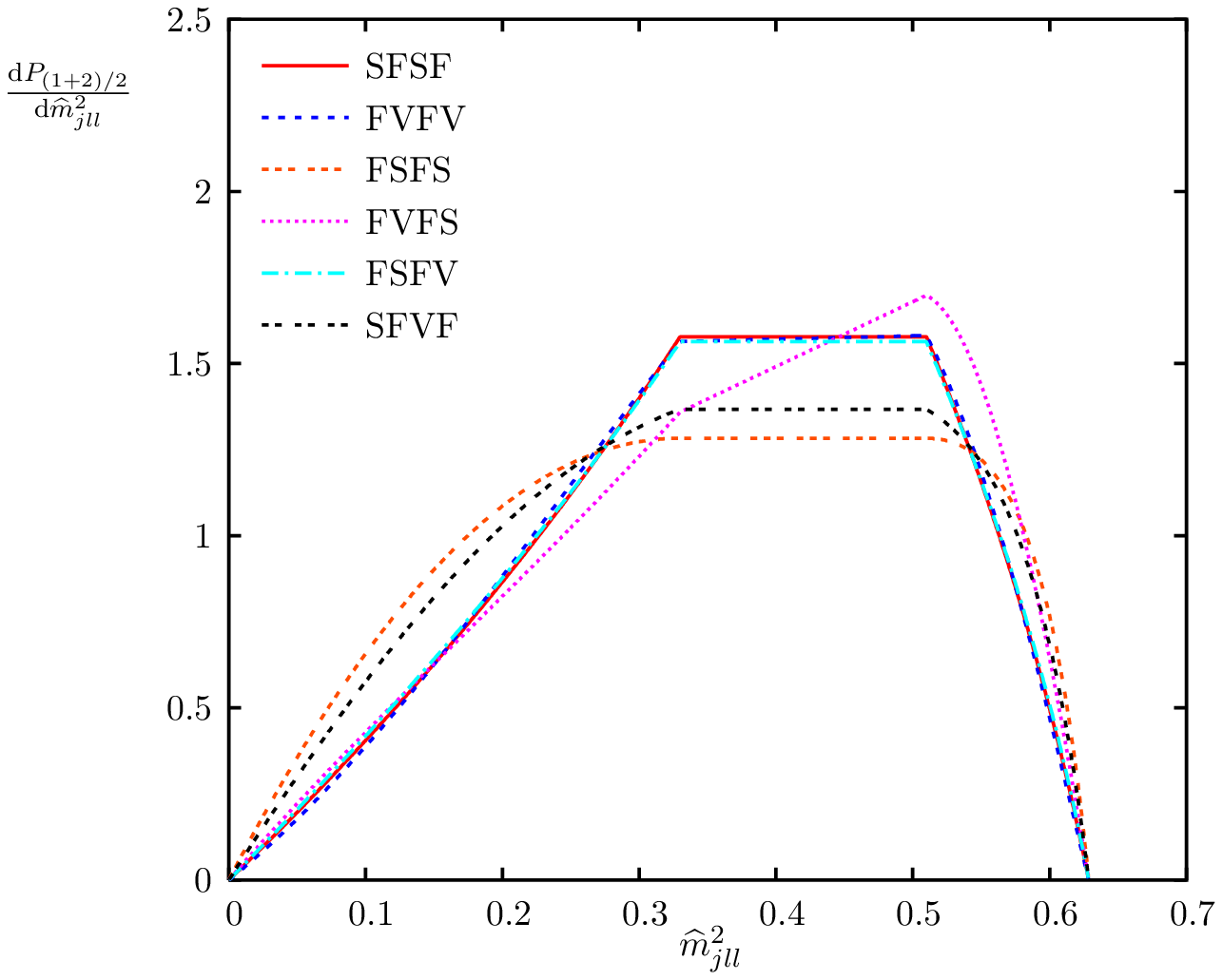}
  \includegraphics[width=0.47\textwidth]{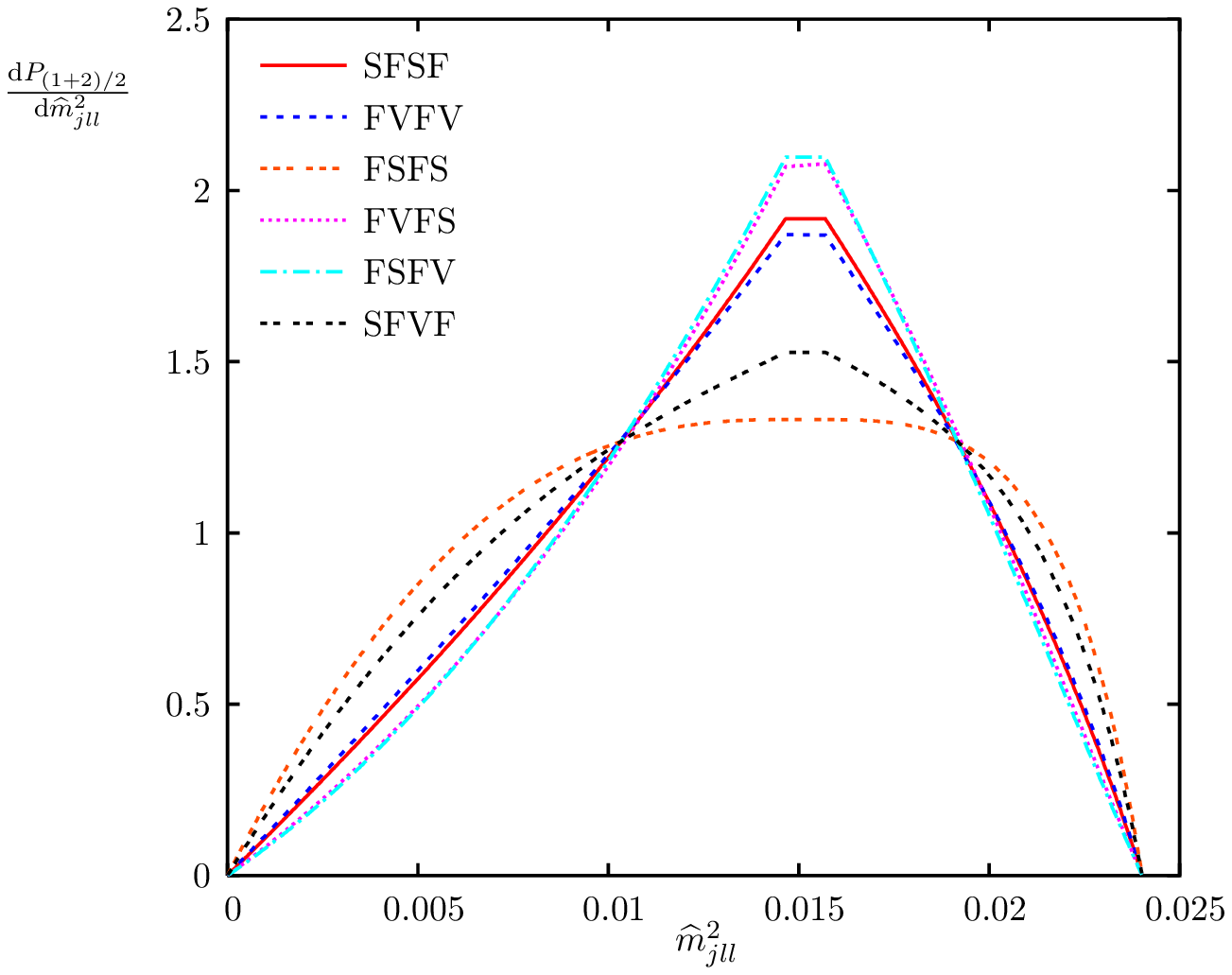}
  \setlength{\unitlength}{1cm}
  \begin{picture}(15,0.5)
    \put(4,0){(a)}
    \put(11.5,0){(b)}
  \end{picture}
  \caption{Jet+dilepton mass distributions for Processes 1 and 2 combined for (a) mass spectrum I
    and (b) mass spectrum II.}
  \label{fig:qllmean}
}

\section{Model discrimination}
\label{sec:model-discrimination}

It is natural to ask whether from {\em real experimental data} it
would be possible to determine the underlying set of particle spins
present in Nature, given only access to one or more of the invariant
mass distributions of section \ref{sec:spin-correlations}.  In particular it would be
useful to know how many events an experiment would need to identify
from one of these decay-chains, in order to favour the assignment of any one
spin configuration over any other at a fixed level of confidence.  We
calculate such a number for each pair of spin configurations and for
each distribution (in isolation from the others) assuming a
``perfect'' detector with infinite acceptance.  As real detectors are
imperfect and will tend to smear the distributions (thereby losing
information) the number of events we calculate should be seen as
representing a {\em lower bound} on the number of events that a real
detector would need to identify.

\par

To calculate the number of events $N$ needed to disfavour a spin
configuration $S$ relative to an alternative spin configuration $T$
(which is assumed to be the actual one generating the observed
events) we solve the following equation for $N$
\begin{equation}
   \frac{1}{R} = \frac{p(S|N{\rm ~events~from~}T)}{p(T|N{\rm ~events~from~}T)},
   \label{eq:thethingwesolveforn}
\end{equation}
in which $R$ is the factor by which configuration $S$ is to be
disfavoured with respect to $T$.  In this note, $R$ was taken to be
1000.  Note that the numerator and the denominator of the
right-hand-side of eq.~(\ref{eq:thethingwesolveforn}) do not
transform into each other under the interchange of $S$ and $T$ as a
result of $T$'s special role as the true configuration chosen in
Nature.  This asymmetry is deliberate, reflecting the fact that there
will only ever be one underlying distribution which generates the
observed events, regardless of the questions which may then be asked
of the nature of those events.\footnote{It would be possible to
write $p(S|N{\rm ~events~from~}U)/p(T|N{\rm ~events~from~}U)$
in place of the right-hand side of eq.~(\ref{eq:thethingwesolveforn})
by introducing a third model $U$ representing the actual production
process.  However, the large number of choices which could be made for $U$
(beyond the already considered choice of $T$) do not suggest that this
course of action would be appropriate in the context of this paper.}

\subsection{One-dimensional analysis}\label{sec:1dKL}
If we characterise the ``$N$ events from $T$'' by the $N$ values of a
particular invariant mass, $m_i^{(T)}$ (for $i\in{1,2,...,N}$),
that are observed in those events, then by using
Bayes' Theorem we can rewrite eq.~(\ref{eq:thethingwesolveforn}) as
\begin{eqnarray}
 \frac{1}{R} &=& 
\frac{p(S)}{p(T)}
\frac
{p(N{\rm ~events~from~}T|S)}
{p(N{\rm ~events~from~}T|T)}
\nonumber\\
&=&
\frac{p(S)}{p(T)}
\frac
{\prod_{i=1}^N p(m_i^{(T)}|S)}
{\prod_{i=1}^N p(m_i^{(T)}|T)}
\nonumber\\
\label{eq:sumgoingtodist}
&=&
\frac{p(S)}{p(T)}
\exp\left(
\sum_{i=1}^N
{\log\left(\frac{p(m_i^{(T)}|S)}{p(m_i^{(T)}|T)}\right)}
\right).
\end{eqnarray}
In the limit $N\gg1$ the sum in eq.~(\ref{eq:sumgoingtodist})
may be approximated by an integral over the allowed masses $m^{(T)}$
weighted according to how prevalent they are:
\begin{eqnarray}
 \frac{1}{R} &\approx&
\frac{p(S)}{p(T)}
\exp\left(
N \int_m
{\log\left(\frac{p(m|S)}{p(m|T)}\right)p(m|T) d m}
\right)
\nonumber\\
&=&\frac{p(S)}{p(T)}
\exp\left(
-N \int_m
{\log\left(\frac{p(m|T)}{p(m|S)}\right)p(m|T) d m}
\right).\label{eq:lumpyfactor}
\end{eqnarray}
The integral in eq.~(\ref{eq:lumpyfactor}) is one that frequently
arises in comparisons of distributions \cite{Kullback:1951} and is
called the {\em Kullback-Leibler distance}\footnote{Note that
the Kullback-Leibler distance is
not symmetric and so does not define a distance in the usual sense. It
is however always non-negative, and is only equal to zero when the two
distributions are identical.} of $S$ from $T$, which we shall denote by:
\begin{equation}\label{eq:KLdist}
{\rm KL}(T,S) = \int_m
{\log\left(\frac{p(m|T)}{p(m|S)}\right)p(m|T) d m}.
\end{equation}
Drawing all these threads together we may rearrange
eq.~(\ref{eq:lumpyfactor}) to give:
\begin{eqnarray}
N \sim \frac{\log R + \log {\frac{p(S)}{p(T)}}}{{\rm KL}(T,S)} \label{eq:nearlyattheend}
\end{eqnarray}
in the limit of large $N$.  Note that the ratio of the prior probabilities for $S$ and $T$
is present in eq.~(\ref{eq:nearlyattheend}) as expected -- strong prior evidence for $S$
over $T$ {\em should} lead to an increase in the number of events $N$ from one of these
chains needed to discredit $S$.  For the numbers presented in tables \ref{tab:AllSUSY.dat}
-- \ref{tab:SqllSUSYmean.dat} and discussed in the following section, however, we assign
equal prior probabilities to $S$ and $T$, thereby removing all prior dependence from
eq.~(\ref{eq:nearlyattheend}).  This choice, in effect, looks at the tests in isolation
from any pre-existing evidence that might favour $S$ over $T$ or vice versa.

Note that the Kullback-Leibler distance is invariant under diffeomorphisms $m \rightarrow
f(m)$ of the distributed variable.  This means in particular that the number of events
$N$ calculated in
eq.~(\ref{eq:nearlyattheend}) does not depend on whether the distributions are
considered to be functions of masses or of masses-squared -- only the intrinsic
information content of the distributions is measured.

\subsection{Three-dimensional analysis}\label{sec:3dKL}
To extract the most information from the data we should compare the predictions
of different spin assigments with the full probability distribution in the
three-dimensional space of $m_{ll}$,  $m_{jl^+}$ and $m_{jl^-}$. The
ambiguity between near and far leptons means that this given by
\beqn\label{P3d}
P(m_{ll},m_{jl^+},m_{jl^-}) &=& \frac 12 f_q \left[P_2(m_{ll},m_{jl^+},m_{jl^-})+P_1(m_{ll},m_{jl^-},m_{jl^+})\right]
\nonumber\\
&+&  \frac 12 f_{\bar q} \left[P_1(m_{ll},m_{jl^+},m_{jl^-})+P_2(m_{ll},m_{jl^-},m_{jl^+})\right]\;,
\eeqn
where we use $P_{1,2}(m_{ll},m_{jl}^{\rm near},m_{jl}^{\rm far})$ on the right-hand
side, assuming both leptons are left-handed, otherwise $f_q$ and $f_{\bar q}$
are interchanged.

Instead of trying to evaluate the three-dimensional generalization of the integral
in eq.~(\ref{eq:KLdist}) analytically, it is convenient to perform a Monte Carlo
integration. If we generate $m_{ll}$, $m_{jl}^{\rm near}$ and $m_{jl}^{\rm far}$
according to phase space, the weight to be assigned to the configuration
$l^{\rm near}=l^+$, $l^{\rm far}=l^-$ is
\beq
P_{+-}(m_{ll},m_{jl}^{\rm near},m_{jl}^{\rm far})
= \frac 12 \left[f_q P_2(m_{ll},m_{jl}^{\rm near},m_{jl}^{\rm far})
           + f_{\bar q} P_1(m_{ll},m_{jl}^{\rm near},m_{jl}^{\rm far})\right]
 \eeq
 while that for $l^{\rm near}=l^-$, $l^{\rm far}=l^+$ is
\beq
P_{-+}(m_{ll},m_{jl}^{\rm near},m_{jl}^{\rm far})
= \frac 12 \left[f_q P_1(m_{ll},m_{jl}^{\rm near},m_{jl}^{\rm far})
           + f_{\bar q} P_2(m_{ll},m_{jl}^{\rm near},m_{jl}^{\rm far})\right]\;.
 \eeq
In the former case, since the distinction between  $l^{\rm near}$ and $l^{\rm far}$ is lost in the data
(except when interchanging them gives a point outside phase space),
we must use eq.~(\ref{P3d}) with $l^+=l^{\rm near}$, $l^-=l^{\rm far}$ in the logarithmic factor of the KL-distance, i.e. the contribution is
\beq
\log\left(\frac{
P_{+-}(m_{ll},m_{jl}^{\rm near},m_{jl}^{\rm far}|T) +
P_{-+}(m_{ll},m_{jl}^{\rm far},m_{jl}^{\rm near}|T)}{
P_{+-}(m_{ll},m_{jl}^{\rm near},m_{jl}^{\rm far}|S) +
P_{-+}(m_{ll},m_{jl}^{\rm far},m_{jl}^{\rm near}|S)}\right)
P_{+-}(m_{ll},m_{jl}^{\rm near},m_{jl}^{\rm far}|T)\;.
\eeq
Similarly from the configuration $l^{\rm near}=l^-$, $l^{\rm far}=l^+$ we get the contribution
\beq
\log\left(\frac{
P_{-+}(m_{ll},m_{jl}^{\rm near},m_{jl}^{\rm far}|T) +
P_{+-}(m_{ll},m_{jl}^{\rm far},m_{jl}^{\rm near}|T)}{
P_{-+}(m_{ll},m_{jl}^{\rm near},m_{jl}^{\rm far}|S) +
P_{+-}(m_{ll},m_{jl}^{\rm far},m_{jl}^{\rm near}|S)}\right)
P_{-+}(m_{ll},m_{jl}^{\rm near},m_{jl}^{\rm far}|T)\;.
\eeq
Denoting the sum of these two contributions at the $i$th phase space point by
${\rm KL}_i(T,S)$, and summing over $M$ such points, we have as $M\to \infty$
\beq
\frac{M\log R}{\sum_i {\rm KL}_i(T,S)}\to N\;,
\eeq
which is the Monte Carlo equivalent of eq.~(\ref{eq:nearlyattheend})
when the prior probabilities of $S$ and $T$ are taken to be equal.
Results for $R=1000$ and $M=5\times 10^7$ are shown in
table~\ref{tab:M3dim.dat} and discussed in the following section.

\section{Discussion and conclusions}
\label{sec:conc}
The results of applying the above one-dimensional analysis to the observable
dilepton, jet+lepton and combined jet+dilepton invariant mass distributions
separately are presented in tables~\ref{tab:AllSUSY.dat}-\ref{tab:SqllSUSYmean.dat},
while the results of the three-dimensional analysis are shown in table~\ref{tab:M3dim.dat}.

\TABLE{
\centering
\begin{tabular}[]{@{}r@{$\,$}|@{$\,$}r@{$\;$}r@{$\;$}r@{$\;$}r@{$\;$}r@{$\;$}r@{}}
 \small (a) $\;$ & \small SFSF & \small FVFV & \small FSFS & \small FVFS & \small FSFV & \small SFVF \\
\hline \small
\small SFSF  & \small $\infty$	& \small 60486	& \small 23	& \small 148	& \small
15608	& \small 66 \\ 
\small FVFV  & \small 60622	& \small $\infty$	& 22	& \small 164	& \small
6866	& \small 62 \\ 
\small FSFS  & \small 36	& \small 34	& \small $\infty$	& 16	& \small
39	& \small 266 \\ 
\small FVFS  & \small 156	& \small 173	& \small 11	& \small $\infty$	&
130	& \small 24 \\ 
\small FSFV  & \small 15600	& \small 6864	& \small 25	& \small 122	& \small
$\infty$	& \small 76 \\ 
\small SFVF  & \small 78	& \small 73	& \small 187	& \small 27	& \small
90	& $\infty$ 
\end{tabular}
\hspace{0.2cm}
\begin{tabular}[]{@{}r@{$\,$}|@{$\,$}r@{$\;$}r@{$\;$}r@{$\;$}r@{$\;$}r@{$\;$}r@{}}
\small (b) $\;$ & \small  SFSF & \small  FVFV & \small  FSFS & \small  FVFS & \small  FSFV & \small  SFVF \\
\hline 
\small SFSF  & \small  $\infty$	& \small  3353	& \small  23	& \small  304	& \small
427	& \small  80 \\ \small  
FVFV  & \small  3361	& \small  $\infty$	& \small  27	& \small  179	& \small
232	& \small  113 \\ \small  
FSFS  & \small  36	& \small  44	& \small  $\infty$	& \small  20	& \small
22	& \small  208 \\ \small  
FVFS  & \small  313	& \small  184	& \small  14	& \small  $\infty$	& \small
13077	& \small  35 \\ \small  
FSFV  & \small  436	& \small  236	& \small  15	& \small  12957	& \small  $\infty$
& \small  39 \\ \small  
SFVF  & \small  89	& \small  126	& \small  134	& \small  38	& \small  42	&
\small  $\infty$ 
\end{tabular}
\caption{The number of events needed to disfavour the column model with respect to the row 
 model by a factor of 0.001, assuming the data to come from the row model, for the
 $\mhat_{ll}^2$ distribution (a) mass spectrum I (figure \ref{fig:ll}a) and (b) mass
 spectrum II (figure \ref{fig:ll}b).}
\label{tab:AllSUSY.dat}
}

\TABLE{
\centering
\begin{tabular}[]{@{}r@{$\,$}|@{$\,$}r@{$\;$}r@{$\;$}r@{$\;$}r@{$\;$}r@{$\;$}r@{}}
 \small (a) $\;$ & \small  SFSF & \small  FVFV & \small  FSFS & \small  FVFS & \small  FSFV & \small  SFVF \\
\hline 
\small SFSF  & \small  $\infty$	& \small  1059	& \small  205	& \small  1524	& \small  758	& \small  727 \\ \small 
FVFV  & \small  1090	& \small  $\infty$	& \small  404	& \small  3256	& \small  4363	& \small  1746 \\ \small 
FSFS  & \small  278	& \small  554	& \small  $\infty$	& \small  418	& \small  741	& \small  870 \\ \small 
FVFS  & \small  1605	& \small  3242	& \small  345	& \small  $\infty$	& \small  1256	& \small  2365 \\ \small 
FSFV  & \small  749	& \small  4207	& \small  507	& \small  1212	& \small  $\infty$	& \small  1803 \\ \small 
SFVF  & \small  813	& \small  1821	& \small  751	& \small  2415	& \small  1888	& \small  $\infty$
\end{tabular}
\hspace{0.2cm}
\begin{tabular}[]{@{}r@{$\,$}|@{$\,$}r@{$\;$}r@{$\;$}r@{$\;$}r@{$\;$}r@{$\;$}r@{}}
\small (b) $\;$ & \small  SFSF & \small  FVFV & \small  FSFS & \small  FVFS & \small  FSFV
& \small  SFVF \\ \hline \small  
SFSF  & \small  $\infty$	& \small  3006	& \small  958	& \small  6874	& \small  761	& \small  1280 \\ \small 
FVFV  & \small  2961	& \small  $\infty$	& \small  4427	& \small  1685	& \small  2749	& \small  3761 \\ \small 
FSFS  & \small  914	& \small  4201	& \small  $\infty$	& \small  743	& \small  9874	& \small  4877 \\ \small 
FVFS  & \small  6716	& \small  1699	& \small  752	& \small  $\infty$	& \small  656	& \small  1306 \\ \small 
FSFV  & \small  720	& \small  2666	& \small  10279	& \small  649	& \small  $\infty$	& \small  4138 \\ \small 
SFVF  & \small  1141	& \small  3517	& \small  5269	& \small  1276	& \small  4259	& \small  $\infty$
\end{tabular}
\caption{As in table \ref{tab:AllSUSY.dat}, for the $\mhat_{jl+}^2$ distribution, (a) mass spectrum I (figure \ref{fig:jlp}a) and (b) mass
 spectrum II (figure \ref{fig:jlp}b).}
\label{tab:KjlpSUSY.dat}
}
\TABLE{
\centering
\begin{tabular}[]{@{}r@{$\,$}|@{$\,$}r@{$\;$}r@{$\;$}r@{$\;$}r@{$\;$}r@{$\;$}r@{}}
 \small (a) $\;$ & \small  SFSF & \small  FVFV & \small  FSFS & \small  FVFS & \small  FSFV & \small  SFVF \\ 
\hline \small 
SFSF  & \small  $\infty$	& \small  1058	& \small  505	& \small  769	& \small  816	& \small  619 \\ \small 
FVFV  & \small  1090	& \small  $\infty$	& \small  541	& \small  5878	& \small  4821	& \small  445 \\ \small 
FSFS  & \small  565	& \small  714	& \small  $\infty$	& \small  1032	& \small  741	& \small  2183 \\ \small 
FVFS  & \small  799	& \small  6435	& \small  882	& \small  $\infty$	& \small  2742	& \small  510 \\ \small 
FSFV  & \small  806	& \small  4641	& \small  507	& \small  2451	& \small  $\infty$	& \small  413 \\ \small 
SFVF  & \small  692	& \small  541	& \small  2272	& \small  576	& \small  521	& \small  $\infty$
\end{tabular}
\hspace{0.2cm}
\begin{tabular}[]{@{}r@{$\,$}|@{$\,$}r@{$\;$}r@{$\;$}r@{$\;$}r@{$\;$}r@{$\;$}r@{}}
 \small (b) $\;$ & \small  SFSF & \small  FVFV & \small  FSFS & \small  FVFS & \small  FSFV & \small  SFVF \\ 
\hline \small 
SFSF  & \small  $\infty$	& \small  3037	& \small  689	& \small  8633	& \small  925	& \small  967 \\ \small 
FVFV  & \small  2985	& \small  $\infty$	& \small  2271	& \small  1431	& \small  4368	& \small  2527 \\ \small 
FSFS  & \small  707	& \small  2297	& \small  $\infty$	& \small  526	& \small  9874	& \small  5004 \\ \small 
FVFS  & \small  8392	& \small  1450	& \small  525	& \small  $\infty$	& \small  653	& \small  843 \\ \small 
FSFV  & \small  924	& \small  4287	& \small  10279	& \small  640	& \small  $\infty$	& \small  4036 \\ \small 
SFVF  & \small  1047	& \small  2693	& \small  5213	& \small  870	& \small  4041	& \small  $\infty$
\end{tabular}
\caption{As in table \ref{tab:AllSUSY.dat}, for the $\mhat_{jl-}^2$ distribution,
  (a) mass spectrum I (figure \ref{fig:jlm}a) and (b) mass
 spectrum II (figure \ref{fig:jlm}b).}
\label{tab:MjlmSUSY.dat}
}
\TABLE{
\centering
\begin{tabular}[]{@{}r@{$\,$}|@{$\,$}r@{$\;$}r@{$\;$}r@{$\;$}r@{$\;$}r@{$\;$}r@{}}
 \small (a) $\;$ & \small  SFSF & \small  FVFV & \small  FSFS & \small  FVFS & \small  FSFV & \small  SFVF \\ \hline \small 
SFSF  & \small  $\infty$	& \small  25630	& \small  241	& \small  1040	& \small  82589	& \small  476 \\ \small 
FVFV  & \small  27315	& \small  $\infty$	& \small  225	& \small  939	& \small  14811	& \small  432 \\ \small 
FSFS  & \small  224	& \small  204	& \small  $\infty$	& \small  265	& \small  252	& \small  2670 \\ \small 
FVFS  & \small  1009	& \small  906	& \small  278	& \small  $\infty$	& \small  1095	& \small  504 \\ \small 
FSFV  & \small  73158	& \small  13688	& \small  269	& \small  1124	& \small  $\infty$	& \small  557 \\ \small 
SFVF  & \small  452	& \small  400	& \small  2749	& \small  493	& \small  533	& \small  $\infty$
\end{tabular}
\hspace{0.2cm}
\begin{tabular}[]{@{}r@{$\,$}|@{$\,$}r@{$\;$}r@{$\;$}r@{$\;$}r@{$\;$}r@{$\;$}r@{}}
 \small (b) $\;$ & \small  SFSF & \small  FVFV & \small  FSFS & \small  FVFS & \small  FSFV & \small  SFVF \\ 
\hline \small 
SFSF  & \small  $\infty$	& \small  26712	& \small  189	& \small  2213	& \small  1686	& \small  421 \\ \small 
FVFV  & \small  26391	& \small  $\infty$	& \small  224	& \small  1323	& \small  1073	& \small  545 \\ \small 
FSFS  & \small  182	& \small  217	& \small  $\infty$	& \small  109	& \small  101	& \small  1710 \\ \small 
FVFS  & \small  2279	& \small  1373	& \small  116	& \small  $\infty$	& \small  46742	& \small  210 \\ \small 
FSFV  & \small  1749	& \small  1121	& \small  109	& \small  47812	& \small  $\infty$
& \small  193 \\ \small 
SFVF  & \small  405	& \small  528	& \small  1712	& \small  196	& \small  179	& \small  $\infty$
\end{tabular}
\caption{As in table \ref{tab:AllSUSY.dat}, for the $\mhat_{jll}^2$ distribution for
  processes 1 and 2 combined,
  (a) mass spectrum I (figure \ref{fig:qllmean}a) and (b) mass
 spectrum II (figure \ref{fig:qllmean}b).}
\label{tab:SqllSUSYmean.dat}
}
\TABLE{
\centering
\begin{tabular}[]{@{}r@{$\,$}|@{$\,$}r@{$\;$}r@{$\;$}r@{$\;$}r@{$\;$}r@{$\;$}r@{}}
 \small (a) $\;$ & \small  SFSF & \small  FVFV & \small  FSFS & \small  FVFS & \small  FSFV & \small  SFVF \\ 
\hline \small 
SFSF  & \small  $\infty$	& \small  455	& \small  21	& \small  47	& \small  348	& \small  55 \\ \small 
FVFV  & \small  474	& \small  $\infty$	& \small  21	& \small  54	& \small  1387	& \small  55 \\ \small 
FSFS  & \small  33	& \small  34	& \small  $\infty$	& \small  13	& \small  39	& \small  188 \\ \small 
FVFS  & \small  55	& \small  67	& \small  10	& \small  $\infty$	& \small  54	& \small  19 \\ \small 
FSFV  & \small  341	& \small  1339	& \small  25	& \small  45	& \small  $\infty$	& \small  66 \\ \small 
SFVF  & \small  62	& \small  64	& \small  143	& \small  19	& \small  79	& \small  $\infty$
\end{tabular}
\hspace{0.2cm}
\begin{tabular}[]{@{}r@{$\,$}|@{$\,$}r@{$\;$}r@{$\;$}r@{$\;$}r@{$\;$}r@{$\;$}r@{}}
 \small (b) $\;$ & \small  SFSF & \small  FVFV & \small  FSFS & \small  FVFS & \small  FSFV & \small  SFVF \\ 
\hline \small 
SFSF  & \small  $\infty$	& \small  1053	& \small  21	& \small  230	& \small  194	& \small  63 \\ \small 
FVFV  & \small  1047	& \small  $\infty$	& \small  27	& \small  135	& \small  190	& \small  90 \\ \small 
FSFS  & \small  33	& \small  42	& \small  $\infty$	& \small  19	& \small  22	& \small  175 \\ \small 
FVFS  & \small  242	& \small  140	& \small  13	& \small  $\infty$	& \small  332	& \small  33 \\ \small 
FSFV  & \small  189	& \small  194	& \small  14	& \small  315	& \small  $\infty$	& \small  37 \\ \small 
SFVF  & \small  66	& \small  95	& \small  118	& \small  35	& \small  41	& \small  $\infty$
\end{tabular}
\caption{As in table \ref{tab:AllSUSY.dat}, for the combined three-dimensional distribution.
(a) mass spectrum I and (b) mass spectrum II.}
\label{tab:M3dim.dat}
}

As expected from figure~\ref{fig:ll}, the dilepton distribution
(table~\ref{tab:AllSUSY.dat})
distinguishes very poorly between the spin assignments SFSF (SUSY)
and FVFV (UED), even with perfect data, requiring over 60k events for
1:1000 discrimination in the case of mass spectrum I.  However, this distribution resolves
well between
FSFS, FVFS, SFVF and other models.\footnote{One should bear
in mind that when only a few events are involved, the approximation of
replacing the sum in eq.~(\ref{eq:sumgoingtodist}) by an integral becomes
invalid, and the discrimination may vary widely between particular data
samples.}

The jet plus lepton distributions (figures~\ref{fig:jlp}, \ref{fig:jlm} and
tables~\ref{tab:KjlpSUSY.dat}, \ref{tab:MjlmSUSY.dat})
are generally more effective than the dileptons
in distinguishing between SFSF and FVFV, but the discrimination is strongly
dependent on the mass spectrum.  As was emphasised in \cite{Smillie:2005ar},
a `quasi-degenerate' spectrum like II, with large values of the mass ratios
$x$, $y$ and $z$ in eq.~(\ref{eq:xyz}), usually makes it more difficult to
resolve between spin assignments. We see however that there are exceptions
to this trend, e.g.\ discrimination between FVFV and FVFS or FSFV.

The jet plus dilepton distribution (figure~\ref{fig:qllmean},
table~\ref{tab:SqllSUSYmean.dat}),
like the dileptons alone, proves ineffective in resolving
between SFSF and FVFV, but discriminates well between FSFS, SFVF
and most other models.

The results in table~\ref{tab:M3dim.dat} for the three-dimensional
analysis show that, as might be expected, this method
achieves a discrimination that is better than that of a one-dimensional
analysis applied to any single invariant mass distribution.  This could be
particularly useful in difficult cases like that of distinguishing between
SFSF (SUSY) and FVFV (UED). 

We should stress again that the numbers of events in the tables correspond to
perfect conditions of signal isolation, resolution and detector efficiency.
Realistic conditions would most likely require much higher numbers.
Nevertheless, the results obtained provide a guide to the places where
attempts to distinguish between spin hypotheses would or would not be
worthwhile. Independent of the merits of the hypotheses considered here,
the Kullback-Leibler distance (\ref{eq:KLdist}) is in our view a useful
tool for addressing questions of this type.

\section*{Acknowledgements}
We thank Sabine Kraml and members of the Cambridge Supersymmetry Working Group for
helpful discussions.

\appendix

\section{Spin correlations from matrix elements}
\label{sec:MEs}
We give here the analytical formulae for the spin correlations in the cascade
decays listed in table~\ref{tab:spins}.
For brevity we write $m_{ql}^{\rm near}$ as $m_{qn}$ and $m_{ql}^{\rm far}$ as
$m_{qf}$ here, and omit propagator denominators and overall normalization
factors. All the distributions presented in this paper are normalized to unity
and so such factors are irrelevant.  SFSF1 denotes scalar-fermion-scalar-fermion
process 1, etc.  The mass ratios $x,y,z$ are defined in eq.~(\ref{eq:xyz}).
\begin{eqnarray}
\mbox{SFSF1:} && \mqn\\
\mbox{SFSF2:} && (1 - x)(1 - y)\mqq - \mqn\\
\mbox{FVFV1:} && x(1 - x)y(1 - y)(1 - z)\mqr - 2x
      (1 - z)[(y - 2xy - 4z + 2yz)\mqq \nonumber\\
      && + 4z\mqn]\mqn - 2x(1 - 2z)[(1 - y)\mqq - 2\mqn]\mqf \nonumber\\
      && + (1 - 2z)[(1 - x)(2 - y)\mqq - 2(1 - 2x)\mqn]\mll\\
\mbox{FVFV2:} && x(1 - x)(1 - y)(1 - z)(y + 8xz)\mqr - 8xz(1 - z)\mqm \nonumber\\
       &&- (1 - 2z) [2x(1 - 2x)(1 - y)\mqf - (1 - x)(2 - y)\mll]\mqq\nonumber\\
       && - 2\{x(1 - z)[y - 2(1 - 2x)(2 - y)z]\mqq - 2x(1 - 2z)\mqf\nonumber\\
       && + (1 - 2z)\mll\}\mqn\\
\mbox{FSFS:} && x(1 - y)(1 - z)\mqq - \mll\\
\mbox{FVFS1:} &&2\{2x\mqf - (1 - 2x)[xy(1 - z)\mqq +
      \mll]\}\mqn \nonumber\\
      && + \{x(1 - y)[(1 - x)y(1 - z)\mqq - 2\mqf] + (1 - x)(2 - y)\mll\}\mqq\\
\mbox{FVFS2:} && -2[xy(1 - z)\mqq - 2x\mqf + \mll] \mqn - 
      \{2x(1 - 2x)(1 - y)\mqf \nonumber\\
      && -(1 - x) [xy(1 - y)(1 - z)\mqq + (2 - y)\mll]\}\mqq\\
\mbox{FSFV:} && x(1 - y)(1 - z)\mqq - (1 - 2z)\mll\\
\mbox{SFVF1:} &&-2x[x(1 - y)(2 - z)\mqq - 2\mll] \mqf
      + 2(1 - x)\{2x^2 y(1 - y)(1 - z)\mqr \nonumber\\
      && + [x(1 - 2y)(2 - z)\mqq - 2\mll]\mll\} + 
      xz[x(1 - 2y)(1 - z)\mqq \nonumber\\
      &&- 2\mll]\mqn\\
\mbox{SFVF2:} && -z[x(1 - 2y)(1 - z)\mqq - 2\mll]\mqn
      + 2[x(1 - y)(2 - z)\mqq - 2\mll]\mqf \nonumber\\
      &&+ (1 - x)z[x(1 - y)(1 - z)\mqq - 2\mll]\mqq
\end{eqnarray}

\section{Analytical formulae for invariant mass distributions}
\label{sec:analytical-formulae}
This section contains all the analytical formulae for the invariant masses for the 6
chains.  Results for the SFSF and FVFV chains have appeared in the literature before
\cite{Smillie:2005ar,Miller:2005zp}, but are included here for completeness.  Throughout,
$m^2$ represents the relevant $\mhat^2$ for that subsection.

\subsection{Dilepton invariant mass distributions}
\label{sec:llPDFS}
The following table (\ref{tab:lldistsc}) contains the dilepton invariant mass
distributions, $\ud P/\ud m_{ll}^2$,
for the different chains.  These are equal in both processes 1 and 2 and have been
normalised to unit area.

\TABLE{
  \centering
  \begin{tabular}{|c|c|}
    \hline Chain & Processes 1 and 2 \\ \hline
    SFSF & $1$ \\ \hline
    FVFV & $\frac{2}{(2+y)(1+2z)} (y+4z+(2-y)(1-2z)m^2)$ \\ \hline
    FSFS & $2(1-m^2)$ \\ \hline
    FVFS & $\frac{2}{2+y}(y+m^2(2-y))$ \\ \hline
    FSFV & $\frac{2}{1+2z}(1-m^2(1-2z))$ \\ \hline
    SFVF & $\frac{3}{(1+2y)(2+z)} (4y+z+4m^2(1-y(2-z)-z)-4m^4(1-y)(1-z))$ \\ \hline
  \end{tabular}
  \caption{Dilepton invariant mass distributions.}
  \label{tab:lldistsc}
}

\subsection{Quark and near lepton invariant mass distributions}
\label{sec:qlnPDFS}

The following tables (\ref{tab:qlnear1} \& \ref{tab:qlnear2}) contain the analytical
forms for the quark and near lepton invariant mass distributions, $\ud P/\ud (m_{ql}^{\rm
  near})^2$.  These have been normalised to unit area.

\TABLE{
  \centering
  \begin{tabular}{|c|c|}
    \hline Chain & Process 1 \\ \hline
    SFSF & $2m^2$ \\ \hline
    FVFV & $\frac{3}{(1+2x)(2+y)} [y+4(1-y+xy)m^2-4(1-x)(1-y)m^4]$ \\ \hline
    FSFS & $1$ \\ \hline
    FVFS & $\frac{3}{(1+2x)(2+y)} [y+4(1-y+xy)m^2-4(1-x)(1-y)m^4]$ \\ \hline
    FSFV & $1$ \\ \hline
    SFVF & $\frac{2(2y+(1-2y)m^2)}{1+2y}$ \\ \hline
  \end{tabular}
  \caption{Quark and near lepton invariant mass distributions for process 1.}
  \label{tab:qlnear1}
}

\TABLE{
  \centering
  \begin{tabular}{|c|c|}
    \hline Chain & Process 2 \\ \hline
    SFSF & $2(1-m^2)$ \\ \hline
    FVFV & $\frac{3}{(1+2x)(2+y)} [4x+y+4(1-2x-y+xy)m^2-4(1-x)(1-y)m^4]$ \\ \hline
    FSFS & $1$ \\ \hline
    FVFS & $\frac{3}{(1+2x)(2+y)} [4x+y+4(1-2x-y+xy)m^2-4(1-x)(1-y)m^4]$ \\ \hline
    FSFV & $1$ \\ \hline
    SFVF & $\frac{2(1+(-1+2y)m^2)}{1+2y}$ \\ \hline
  \end{tabular}
  \caption{Quark and near lepton invariant mass distributions for process 2.}
  \label{tab:qlnear2}
}
It is to be expected that the FVFV and FVFS distributions and the FSFS and FSFV
distributions match as the chains are identical up to the first two vertices in the chain
and the quark and near lepton invariant mass is unaffected by what happens at the third
vertex.

\subsection{Quark and far lepton invariant mass distributions}
\label{sec:qlfPDFS}

The following gives the quark and far lepton invariant mass distributions for the
different processes, $\ud P/\ud (m_{ql}^{\rm far})^2$, normalised to unit area.  The index
1 or 2 indicates Process 1 or 2 respectively.

\textbf{SFSF}
\begin{equation}
  \label{eq:SFSFqlf1}
  \frac{\ud P_1}{\ud m^2} = \frac{-2}{(1-y)^2} \left\{
    \begin{array}{l @{\quad} c}
      (1-y+\log y) & 0 \le m^2 \le y \\ & \\
      (1-m^2+\log m^2) & y \le m^2 \le 1 \\
    \end{array} \right.
\end{equation}
\begin{equation}
  \label{eq:SFSFqlf2}
  \frac{\ud P_2}{\ud m^2} = \frac{2}{(1-y)^2} \left\{
    \begin{array}{l @{\quad} c}
      (1-y+y\log y) & 0 \le m^2 \le y \\ & \\
      (1-m^2+y\log m^2) & y \le m^2 \le 1 \\
    \end{array} \right.
\end{equation}

\textbf{FVFV}
\begin{eqnarray}
  \label{eq:FVFVqlf1}
  \frac{\ud P_1}{\ud m^2} &=&\frac{6}{(1+2x)(2+y)(1+2z)(1-y)^2} \times \nonumber \\ &&
  \left\{ 
      \begin{array}{l @{\quad} c}
        (1-y)[4x-y+2(2+3y-2x(5+y))z & \\
        \quad -4m^2(2-3x)(1-2z)]-[y(1-2(4+y)z)+4x(2z-y(1-4z))& \\ \qquad+4m^2(1+y-x(2+y))(1-2z)]\log y & 0
        \le m^2 \le y \\ & \\
        (1-m^2)[4x(1+2y-5z-6yz)-5y+2(2+9y)z & \\
        \quad -4m^2(1-x)(1-z)]-[y(1-2(4+y)z)+4x(2z-y(1-4z)) & \\ 
        \qquad +4m^2(1+y-x(2+y))(1-2z)] \log m^2 & y \le m^2 \le 1
      \end{array} \right.
\end{eqnarray}
\begin{eqnarray}
  \label{eq:FVFVqlf2}
  \frac{\ud P_2}{\ud m^2} &=& \frac{6}{(1+2x)(2+y)(1+2z)(1-y)^2} \times \nonumber \\ &&
  \left\{
    \begin{array}{l @{\quad} c}
      (1-y)[-y+2(2+2x(1-y)+3y)z-4m^2(2-x)(1-2z)] & \\
      \quad -[y(1-2(4+y)z)+4m^2(1+(1-x)y)(1-2z)] \log y & 0 \le m^2 \le y \\ & \\
      (1-m^2)[4(1+x)z-y(5-18z+8xz)-4m^2(1-x)(1-z)] & \\
      \quad -[y(1-2(4+y)z)+4m^2(1+(1-x)y)(1-2z)]\log m^2 & y \le m^2 \le 1 \\
    \end{array} \right.
\end{eqnarray}

\textbf{FSFS}
\begin{eqnarray}
  \label{eq:FSFSqlf12}
  \frac{\ud P_{1,2}}{\ud m^2} = \frac{-2}{(1-y)^2} \left\{
    \begin{array}{l @{\quad} c}
      (1-y+\log y) & 0\le m^2 \le y \\ & \\
      (1-m^2+ \log m^2) & y \le m^2 \le 1 \\
    \end{array} \right.
\end{eqnarray}

\textbf{FVFS}
\begin{eqnarray}
  \label{eq:FVFSqlf1}
  \frac{\ud P_1}{\ud m^2}=\frac{6}{(1+2x)(2+y)(1-y)^2} \left\{
    \begin{array}{l @{\quad} c}
      (1-y)(4x-y-4m^2(2-3x))& \\ \quad +[(-1+4x)y& \\ \quad \quad +4m^2(2x-1-(1-x)y)]\log y & 0\le m^2 \le
      y \\ & \\
      (1-m^2)(4x-5y+8xy-4m^2(1-x))& \\ \quad +[(-1+4x)y& \\ \quad \quad +4m^2(2x-1-(1-x)y)] \log m^2 & y \le m^2
      \le 1 \\
    \end{array} \right. 
\end{eqnarray}
\begin{eqnarray}
  \label{eq:FVFSqlf2}
  \frac{\ud P_2}{\ud m^2}=\frac{6}{(1+2x)(2+y)(1-y)^2} \left\{
    \begin{array}{l @{\quad} c}
      (1-y)(-y-4m^2(2-x)) & \\ \quad -(y+4m^2(1+(1-x)y))\log y & 0 \le m^2 \le y \\ & \\
      (1-m^2)(-5y-4m^2(1-x)) & \\ \quad -(y+4m^2(1+(1-x)y))\log m^2 & y \le m^2 \le 1
    \end{array} \right. 
\end{eqnarray}

\textbf{FSFV}
\begin{eqnarray}
  \label{eq:FSFVqlf}
  \frac{\ud P_{1,2}}{\ud m^2} = \frac{-2}{(1+2z)(1-y)^2} \left\{ 
    \begin{array}{l @{\quad} c}
      (1-y)(1-2z)+(1-2yz)\log y & 0 \le m^2 \le y \\ & \\
      (1-m^2)(1-2z)+(1-2yz) \log m^2 & y \le m^2 \le 1
    \end{array} \right.
\end{eqnarray}

\textbf{SFVF}
\begin{equation}
  \label{eq:SFVFqlf1}
  \frac{\ud P_1}{\ud m^2} = \frac{6}{(1+2y)(2+z)(1-y)^2} \left\{ 
    \begin{array}{l @{\quad} c}
      (1-y)(2-3z-2y(1+z)+4m^2(1-2z)) & \\
      \quad -(z+4yz-4m^2(1-z-yz))\log y & 0 \le m^2 \le y \\ & \\
      (1-m^2)(2-3z-8yz+2m^2(1-z)) & \\
      \quad -(z+4yz-4m^2(1-z-yz))\log m^2 & y \le m^2 \le 1 \\
    \end{array} \right.
\end{equation}

\begin{equation}
  \label{eq:SFVFqlf2}
  \frac{\ud P_2}{\ud m^2} = \frac{6}{(1+2y)(2+z)(1-y)^2} \left\{
    \begin{array}{l @{\quad} c}
      (1-y)(z-4m^2(1-2z)) & \\ \quad +(yz-4m^2(1-z-yz))\log y & 0 \le m^2 \le y \\ & \\
      (1-m^2)(z(1+4y)-4m^2(1-z)) & \\ \quad +(yz-4m^2(1-z-yz)) \log m^2 & y \le m^2
      \le 1 \\
    \end{array} \right.
\end{equation}

\subsection{Quark, near and far lepton mass distributions}
\label{sec:qllPDFS}
Due to the complicated nature of the $qll$ mass, these distributions can be lengthy and
complicated.  Included in this section are the ones of manageable length - FSFS, FSFV and
SFSF (processes 1 and 2).  The others are available from the authors on request.

These distributions are given in terms of 
\begin{equation}
  \label{eq:MllPlus}
  \mathcal{M}_{ll}^+ = \mathrm{min}\left\{
    \begin{array}{c}
      \frac{ \frac{1}{4} (1-x) \left(\sqrt{(1-\sqrt{xyz})^2 - \hat{m}_{qll}^2}
        + \sqrt{(1+\sqrt{xyz})^2 - \hat{m}_{qll}^2}\right)^2-(1-x-\hat{m}_{qll}^2)}{x(1-y)(1-z)}
      \\ \\
      1.
    \end{array}
    \right.
\end{equation}

This is equal to 1 in the region between $(1-y)(1-xz)$ and $(1-z)(1-xy)$, which leads to
the flat section of the distributions observed.  These distributions require the positive
constants $\mathcal{N}_i$ to be set such that the distribution integrates to 1 in each case.

\textbf{FSFS}
\begin{eqnarray}
  \label{eq:qllFSFS}
  \frac{\ud P}{\ud \hat{m}_{qll}^2} &=& \frac{\mathcal{N}_1}{(1-y)^2(1-z)^2} \Bigg\{
     \\ && \quad (1-yz)-\sqrt{((1+\sqrt{yz})^2-(1-y)(1-z)
    \llp) ((1-\sqrt{yz})^2-(1-y)(1-z) 
      \llp)} \nonumber \\ &&  
    -2(y+z)\log{\left(\frac{\sqrt{(1+\sqrt{yz})^2-(1-y)(1-z)
        \llp}-\sqrt{(1-\sqrt{yz})^2-(1-y)(1-z) \llp}}{2\sqrt{yz}}\right)} \Bigg\}\nonumber. 
\end{eqnarray}

\textbf{FSFV}
\begin{eqnarray}
  \label{eq:qllFSFV}
  \frac{\ud P}{\ud \hat{m}_{qll}^2} &=&  \frac{\mathcal{N}_2}{(1-y)^2(1-z)^2} \Bigg\{
     (1-2z)(1-yz) \\ && \quad -(1-2z)\sqrt{((1+\sqrt{yz})^2-(1-y)(1-z)
    \llp) ((1-\sqrt{yz})^2-(1-y)(1-z) 
      \llp)} \nonumber \\ &&  \quad
    -2(y+z-2z(1+yz)) \times \nonumber \\ && \quad
    \log{\left(\frac{\sqrt{(1+\sqrt{yz})^2-(1-y)(1-z)
        \llp}-\sqrt{(1-\sqrt{yz})^2-(1-y)(1-z) \llp}}{2\sqrt{yz}}\right)} \Bigg\}\nonumber. 
\end{eqnarray}

\textbf{SFSF}
\beqn
\label{eq:qllSFSF1}
\frac{\ud P_1}{\ud \hat{m}_{qll}^2} &=& \frac{\mathcal{-N}_3}{(1-y)^2(1-x)(1-z)} \Bigg\{
\frac{1}{z} ((1+xz)(1-yz)-\hat{m}_{qll}^2(1-z)) \\ && \quad
+ \frac{\llp(1-y)(1-z)((1+xz)(1+yz) + 2z(1+xy) - \hat{m}_{qll}^2(1+z))}{z
  \sqrt{(1-\sqrt{yz})^2-\llp(1-y)(1-z)}
  \sqrt{(1+\sqrt{yz})^2-\llp(1-y)(1-z)}} \nonumber \\ && \quad
+\frac{(1-yz)(\hat{m}_{qll}^2(1-z) - (1+xz)(1-yz))}{z
  \sqrt{(1-\sqrt{yz})^2-\llp(1-y)(1-z)} \sqrt{(1+\sqrt{yz})^2-\llp(1-y)(1-z)}} \nonumber
\\ && \quad -4(1+xy) \log{\left(\frac{\sqrt{(1+\sqrt{yz})^2-(1-y)(1-z)
        \llp}-\sqrt{(1-\sqrt{yz})^2-(1-y)(1-z) \llp}}{2\sqrt{yz}}\right)} \Bigg\}\nonumber.
\eeqn

\beqn
\label{eq:qllSFSF2}
\frac{\ud P_2}{\ud \hat{m}_{qll}^2} &=& 
-\frac{\ud P_1}{\ud \hat{m}_{qll}^2}  \\ && 
 +\frac{8 \mathcal{N}_3}{(1-y)(1-z)} 
\log{\left(\frac{\sqrt{(1+\sqrt{yz})^2-(1-y)(1-z) \llp}-\sqrt{(1-\sqrt{yz})^2-(1-y)(1-z)
        \llp}}{2\sqrt{yz}}\right)} \Bigg\} \nonumber
\eeqn

An explicitly normalised expression for an equal 
mixture of process 1 and process 2 for the SFSF (supersymmetric) 
spin configuration may be found in \cite{Lester:2006yw}.


\bibliography{Refs}
\bibliographystyle{utphys}

\end{document}